\documentclass[longauth]{aa}  
\usepackage{color,soul}
\usepackage{natbib}
\usepackage{subfigure}
\usepackage{subfig}
\usepackage{verbatim}
\usepackage{graphicx}
\usepackage{txfonts}
\begin{document}

\title{Lower-than-expected flare temperatures for TRAPPIST-1}

   \author{A. J. Maas
          \inst{\ref{inst:UH},}\inst{\ref{inst:UKM},}\inst{\ref{inst:LSW}}
          \and E. Ilin \inst{\ref{inst:AIP},}\inst{\ref{inst:UP}} 
          \and M. Oshagh \inst{\ref{inst:IAC},}\inst{\ref{inst:ULL}}
          \and E. Pallé \inst{\ref{inst:IAC},}\inst{\ref{inst:ULL}}
          \and H. Parviainen \inst{\ref{inst:IAC},}\inst{\ref{inst:ULL}}
          \and K. ~Molaverdikhani \inst{\ref{inst:USM},}\inst{\ref{inst:LSW},}\inst{\ref{inst:Origins}}
          \and A. ~Quirrenbach     \inst{\ref{inst:UH},}\inst{\ref{inst:LSW}}
          \and E.~Esparza-Borges \inst{\ref{inst:IAC},}\inst{\ref{inst:ULL}}
          \and F. Murgas \inst{\ref{inst:IAC},}\inst{\ref{inst:ULL}}
          \and V. J. S. Béjar \inst{\ref{inst:IAC},}\inst{\ref{inst:ULL}}
          \and N. Narita \inst{\ref{inst:IAC},}\inst{\ref{inst:AC},}\inst{\ref{inst:NAO}}
          \and A. Fukui  \inst{\ref{inst:AC},}\inst{\ref{inst:IFS}} 
          \and C.-L. Lin \inst{\ref{inst:GIA}}
          \and M. Mori  \inst{\ref{inst:DAT}}
          \and P. Klagyivik \inst{\ref{inst:DLR}} 
          \fnmsep\thanks{Supporting Material: The photometry of the two flares in $g$,$r$,$i$, and $z_\mathrm{s}$ filters is available online at the CDS via anonymous ftp to cdsarc.u-strasbg.fr (130.79.128.5). }}

   \institute{Department for Physics and Astronomy University of Heidelberg, Im Neuenheimer Feld 226, D-69120 Heidelberg, Germany  \label{inst:UH}\\ 
              \email{aaron.maas@uni-heidelberg.de} 
         \and Department for Medicine of Georg-August University Göttingen, Robert-Koch-Straße 40, D-37075 Göttingen , Germany \label{inst:UKM} 
         \and Instituto de Astrofísica de Canarias \\
             Calle Vía Láctea, s/n, 38205 San Cristóbal de La Laguna, Santa Cruz de Tenerife, Spain \label{inst:IAC}\\
             \email{epalle@iac.de}
        \and Deptartamento de Astrofísica Universidad de La Laguna (ULL) \\ 
            E-38206 La Laguna, Tenerife, Spain \label{inst:ULL}
        \and Leibniz Insitut für Astrophysik Potsdam (AIP) \\
            An der Sternwarte 16, 14482 Potsdam, Germany \label{inst:AIP} \\
            \email{eilin@aip.de}
        \and Institute for Physics and Astronomy, University of Potsdam \\
            Karl-Liebknecht-Str. 24/25, 14476 Potsdam, Germany \label{inst:UP}
        \and Landesternwarte Königstuhl (LSW),\\
                 Zentrum für Astronomie der Universität Heidelberg, Königstuhl 12, D-69117 Heidelberg, Germany \label{inst:LSW}
        \and Universitäts-Sternwarte (USM),  \\       Ludwig-Maximilians-Universität München, Scheinerstrasse 1, D-81679 München, Germany  \label{inst:USM}   
        \and ORIGINS, Exzellenzcluster Origins, \\ Boltzmannstraße 2, 85748 Garching, Germany \label{inst:Origins}
        \and Okayama Astrophysical Observatory, \\ 
            National Astronomical Observatory of Japan, Asakuchi, Okayama 719-0232, Japan \label{inst:Okayama}
        \and Astrobiology Center, \\
             2-21-1 Osawa, Mitaka, Tokyo, 181-8588, Japan \label{inst:AC}
        \and National Astronomical Observatory of Japan, \\
             2-21-1 Osawa, Mitaka, Tokyo 181-8588, Japan  \label{inst:NAO}
        \and Komaba Institute for Science, The University of Tokyo, 3-8-1 Komaba, Meguro, Tokyo 153-8902, Japan \label{inst:IFS}
        \and Graduate Institute of Astronomy, National Central University, Taoyuan 32001, Taiwan \label{inst:GIA}
        \and Department of Astronomy,  Graduate School of Science, The University of Tokyo, 7-3-1 Hongo, Bunkyo, Tokyo 113-0033, Japan \label{inst:DAT}
        \and Freie Universität Berlin, Institute of Geological Sciences, Malteserstr. 74-100, D-12249 Berlin, Germany \label{inst:DLR}
        }

%Astrobiology Center, 2-21-1 Osawa, Mitaka, Tokyo 181-8588, Japan

    %date received 
   \date{Received April 26, 2022; accepted August 26, 2022}

% \abstract{}{}{}{}{} 
% 5 {} token are mandatory
 
  % \abstract{}{}{}{}{} 
% 5 {} token are mandatory
  
  \abstract
   {}
   { Stellar flares emit thermal and nonthermal radiation in the X-ray and ultraviolet (UV) regime. Although high energetic radiation from flares is a potential threat to exoplanet atmospheres and may lead to surface sterilization, it might also provide the extra energy for low-mass stars needed to trigger and sustain prebiotic chemistry. Despite the UV continuum emission being constrained partly by the flare temperature, few efforts have been made to determine the flare temperature for ultra-cool M-dwarfs. We investigate two flares on TRAPPIST-1, an ultra-cool dwarf star that hosts seven exoplanets of which three lie within its habitable zone. The flares are detected in all four passbands of the MuSCAT2 instrument allowing a determination of their temperatures and bolometric energies.} 
   {We analyzed the light curves of the MuSCAT1 (multicolor simultaneous camera for studying atmospheres of transiting exoplanets) and MuSCAT2 instruments obtained between 2016 and 2021 in $g,r,i,z_\mathrm{s}$-filters. We conducted an automated flare search and visually confirmed possible flare events. The black body temperatures were inferred directly from the spectral energy distribution (SED) by extrapolating the filter-specific flux. We studied the temperature evolution, the global temperature, and the peak temperature of both flares.}
   {White-light M-dwarf flares are frequently described in the literature by a black body with a temperature of 9000-10000 K. For the first time we infer effective black body temperatures of flares that occurred on TRAPPIST-1. The black body temperatures for the two TRAPPIST-1 flares derived from the SED are consistent with $ T_\mathrm{SED} = 7940_{-390}^{+430} $ K and $ T_\mathrm{SED} = 6030_{-270}^{+300} $ K. The flare black body temperatures at the peak are also calculated from the peak SED yielding $T_\mathrm{SEDp} =  13620_{-1220}^{1520}$ K and $ T_\mathrm{SEDp} =  8290_{-550}^{+660}$ K. We update the flare frequency distribution of TRAPPIST-1 and discuss the impacts of lower black body temperatures on exoplanet habitability.} 
   {We show that for the ultra-cool M-dwarf TRAPPIST-1 the flare black body temperatures associated with the total continuum emission are lower and not consistent with the usually adopted assumption of 9000-10000 K in the context of exoplanet research. For the peak emission, both flares seem to be consistent with the typical range from 9000 to 14000 K, respectively. This could imply different and faster cooling mechanisms. Further multi-color observations are needed to investigate whether or not our observations are a general characteristic of ultra-cool M-dwarfs. This would have significant implications for the habitability of exoplanets around these stars because the UV surface flux is likely to be overestimated by the models with higher flare temperatures. } 
   
   \maketitle
%
%-------------------------------------------------------------------

\section{Introduction}
\label{sec:Introduction}

M-dwarfs are prime targets for the search for habitable conditions, that is, rocky planets in the habitable zone of the stars. The small stellar radius and low effective temperatures of M-dwarfs permit not only the detection of habitable planets but also the atmospheric characterization of those in close-in orbits \citep[e.g.][and references therein]{2009_Kaltenegger}. M-dwarfs represent roughly $~73 \%$ of all stars \citep{1964_Dole, 2006_Henry, 2018_Henry}. Late M-dwarfs, including ultra-cool M-dwarfs, have masses $M_* \leq 0.2 M_\odot$ and effective temperatures $T_\mathrm{eff} \leq 3000$ K \citep{2017_Gillon}. Ultra-cool dwarfs are stellar and substellar objects with spectral type later than M7V \citep{1995_Kirkpatrick,1997_Kirkpatrick}. Mid and early M-dwarfs are more massive, $ 0.2 M_\odot < M_* \leq 0.4 M_\odot$, $0.4 M_\odot < M_* \leq 0.6 M_\odot$ respectively, and thus have surface temperatures of up to 3900 K \citep{1999_Heath,2006_Farihi}. Ground-based targeted searches such as MEarth \citep{2009_Irwin}, TRAPPIST \citep{Gillon_2013}, ExTrA \citep{2015_Binfils} and SPECULOOS \citep{Sebastian_2020}, as well as the space missions Kepler \citep{Koch_2010}, K2 \citep{Howell_2014}, and  TESS \citep{Ricker_2014}, have boosted the discoveries of exoplanets around nearby M-dwarfs. Consequently, there is growing interest in putting constraints on the flare rates of the host stars and the respective flare energies, temperatures, and areas  \citep[e.g][]{2008_Fuhrmeister, 2016_Schmidt, Davenport_2016,2017_Vida, Gunther_2020, Howard_2020, 2021_Johnson} in order to understand how these can affect planetary atmospheres and the prospects of searching for habitable conditions.

%A flare is considered a super-flare, if it reaches bolometric energies of $10^{33} - 10^{38}$ erg, \citep{2013_Shibayama}. Solar flares range usually between $10^{29} - 10^{33}$ erg, e.g \citep{2002_Shibata}. The flare-event with the biggest recorded energy release of the sun was the "Carrington Event" in 1895 releasing an energy of $10^{32}$ erg, \citep{Carrington_1863}. Energetic flare events like the Carrington event are often accompanied by Coronal Mass Ejections (CME's), i.e. isotropic high energetic particle emission. 

Stellar flares are explosive magnetic reconnection events \citep{1989_Petterson} occurring in main sequence (MS) stars with convective envelopes. When opposing magnetic field lines approach each other, they are prone to reconnect and release the energy that was stored in the magnetic field in the form of accelerated particles and electromagnetic radiation. These stochastic events have a duration ranging from minutes to several hours \citep{Benz_2017}[e.g.][and references therein]. 
The flare occurrence frequency and the released energy are both dependent on the magnetic field of the stellar surface, which is known to decrease with stellar age for MS stars \citep{2022_Reiners}. Due to the loss of angular momentum, the stellar dynamo quiets during the star's lifetime  \citep{1972_Skumanich}. For low-mass stars, this process takes longer than for solar type stars and thus, in comparison, low-mass stars experience a prolonged active time \citep{2008_West}, they produce high UV and X-ray flux over longer periods \citep{2000_Audard,2014_Hawley}, which can alter the atmospheres of close-by planets and might be a serious threat to surface life \citep{2010_segura, 2019_Tilley}. Other studies suggest that additional UV flux from stellar flares might trigger and drive prebiotic chemistry on planetary surfaces \citep{2017_Ranjan,Rimmer_2018}. Several recent studies of stellar flares on nearby M-dwarfs with planetary companions address surface habitability \citep{Ribas_2016, 2017_OmalleyJames, Meadows_2018,Vida_2019,Glazier_2020} and use constant flare temperatures of $T_\mathrm{eff} = 9000-10000$ K. However, a lower or higher flare temperature would impact the UV flux models for the star and therefore the flux incident on the planetary companions.  %Thus, the most of the recent studies addressing surface habitability around late Ms are effected by our results. 

The primary output of flares are accelerated particles, which then precipitate back on the chromosphere of the star, emitting nonthermal radiation and subsequently heat the chromospheric plasma \citep{2010_Benz, 2017_Aschwanden, Benz_2017}. The amount of heating differs locally and is strongest at the footpoints of the magnetic loop after reconnection. This local heating gives rise to thermal continuum emission and line emission in the optical, UV, and soft X-ray regime \citep{2010_Benz, 2013_Kowalski, 2017_Aschwanden}. One example is the H$\alpha$ line. At the footpoints of the magnetic fields, accelerated electrons and protons will recombine producing hydrogen emission lines, the so-called H$\alpha$ ribbons \citep{2013_Kowalski}. The main contribution to the total energy budget of a flare comes from the white-light continuum emission \citep{2010_Benz,2017_Aschwanden}. The total energy budget is frequently estimated by a black body of effective temperature $T_\mathrm{eff} = 9000-10000$ K  \citep{1991_Hawley,2011_Kretzschmar, 2013_Kowalski, 2013_Shibayama, Davenport_2016, Paudel_2018, 2018_Howard, 2018_Jackman, 2019_Jackman,  Gunther_2020} . While this knowledge relies mainly on magnetically active M-dwarfs (dMe stars) with spectral types dM3e-dM4.5e from spectroscopic and photometric observations\citep[e.g.][and references therein]{2013_Kowalski}, it is usually assumed that flares on ultra-cool M-dwarfs like TRAPPIST-1 with spectral type M7.5V behave similarly \citep{Paudel_2018, Gunther_2020, Glazier_2020}. 
% important not persitent chromosphere!!!!!

%g. Hawley & Pettersen 1991; Kowalski et al. 2013;
%Shibayama et al. 2013; Davenport 2016; Jackman et al.
%2018, 2019; Howard et al. 2018; Chang et al. 2018

%Potentially dangerous radiation to planetary atmospheres like UV and X-ray can be estimated from white-light observation by making a black body assumption. However, it is obvious without considering the nonthermal contribution from the accelerated particles there will be uncertainty on the  amount of emitted high energetic flux, e.g. \cite{Howard_2020}

Multi-passband photometry is a useful tool for determining continuum black body temperatures of stellar flares \citep{Hawley_2003, 2013_Kowalski, Howard_2020}. The black body temperature of stellar flares is not constant and should be understood as an energy budget of the continuum emission. As flares are subject to different physical processes in different atmospheric layers, it is nontrivial to attribute a single temperature characterizing the flare. If flares are observed in several photometric filters, recent methods can be employed to retrieve black body temperatures \citep{Howard_2020}. For typical active M-dwarf stars spectroscopic and photometric observations are consistent with a black body with temperatures of 9000-14000 K at the peak and 5500-7000 K in the gradual phase \citep{2013_Kowalski}.  \cite{Howard_2020} investigated a large sample of stars with photometric flares and spectral types of M0-M7V. These authors found the total temperature averaged over their sample to be $T_\mathrm{eff} = 11000 \substack{+3500\\-2600}$ K and at peak $T_\mathrm{eff} = 14000 \substack{+8300\\-3400}$ K.

% Solar flares range usually between $10^{29} - 10^{33}$ erg, e.g \citep{2002_Shibata}. The flare-event with the biggest recorded energy release of the sun was the "Carrington Event" in 1895 releasing an energy of $10^{32}$ erg, \citep{Carrington_1863}. Energetic flare events like the Carrington event are often accompanied by Coronal Mass Ejections (CME's), i.e. isotropic high energetic particle emission.

For studying flares on low-mass stars in the context of exoplanet
research, the most promising target is TRAPPIST-1. This target is a well-studied late M-dwarf with spectral type M7.5V \citep{2000_Gizis} and is known to host seven rocky exoplanets of which three lie within the habitable zone, which is the zone where the surface flux is just right for water to be liquid
\citep{2016_Gillom, 2017_Gillon, 2017_Luger}. Because TRAPPIST-1 flares frequently, with a flare with an energy above $10^{29}$ erg occurring
every 1-2 days  \citep{Paudel_2018}, the impact of its flaring activity 
on the atmospheres of its orbiting exoplanets is being investigated  
\citep{2017_Vida,2017_Malley,Paudel_2018,Glazier_2020,2020_Estrela}. 
\cite{Paudel_2018} analysed K2 data for TRAPPIST-1 and detected 39 flares, 
and a superflare rate of  $4.4\substack{+2.0 // -0.2}$ flares per year. A flare 
is considered a superflare if it reaches bolometric energies of $10^{33} 
- 10^{38}$ erg \citep{2013_Shibayama}. \cite{Glazier_2020} did not detect
flares in a two-year survey with Everyscope, an array of small telescopes
imaging the entire accessible sky all at once \citep{2016_law, 2019_Ratzloff}. \cite{Glazier_2020} and \cite{Paudel_2018} argue
that the possible atmospheres of the planets in the habitable zone of
TRAPPIST-1 are, with the current flare rate, not in danger of complete
ozone depletion, but also suggest that the UV flux from TRAPPIST-1 is not enough to trigger and sustain abiogenesis. \cite{2019_Tilley} showed that
UV radiation and proton fluxes from frequent flaring can deplete the ozone layer of an Earth-like planet almost entirely over timescales of
several million years. \cite{2020_Estrela} examined the impact of the UV radiation from flares on the potentially habitable planets of TRAPPIST-1.
These authors argue that UV-sensitive organisms could survive if the planet has an ozone layer or if their natural living environment is at least 8m
below surface ocean level. These studies highlight the
importance of considering flares when assessing the habitability of a planet. Despite the growing interest in flaring processes of ultra-cool M-dwarfs and their subsequent UV radiation, the flare temperature of the continuum emission remain uncertain for late M-dwarfs. For a L-dwarf (L1V) flare, \cite{Gizis_2013}found a black body spectrum consistent with an effective temperature of $T_\mathrm{eff}= 8000 \pm 2000$ K. 

Here, we present an automated search for stellar flares on TRAPPIST-1 using MuSCAT1 (Multicolor simultaneous camera for studying atmospheres of transiting exoplanets) \citep{Narita_2015} and MuSCAT2 \citep{Narita_2019} data and the open source package \textit{AltaiPony} \citep{Ilin_2021b}. For the first time using the simultaneous multi-color photometry of the MuSCAT instruments, we estimated effective black body temperatures of TRAPPIST-1 flares. 

The paper is structured as follows: In Section \ref{sec:Observation}, we present the instruments, observations and preparation of the photometric data. Section \ref{sec:flaredection} describes the flare detection, and in Section \ref{sec:Methods}, we describe the methods used to derive black body temperatures. The results are presented in Section \ref{sec:Results} and discussed in Section \ref{sec:Discussion}. Finally, we summarize our findings in Section \ref{sec:Summary}.

\section{Observations}
\label{sec:Observation}

%We analyse photometric data from MuSCAT1 (M1), MuSCAT2 (M2) and Kepler-K2 obtained between 2016-2020. 

MuSCAT1 (M1) and MuSCAT2 (M2) (Multicolor Simultaneous Camera for studying Atmospheres of Transiting exoplanets) are part of the Global Multi-Color Photometric Monitoring Network for Exoplanetary Transits. M1 is mounted on the 1.88m telescope at the Okayama Astro-Complex in Japan and M2 is located at the Teide Observatory in Tenerife, Spain and mounted on the 1.52 m Telescopio Carlos Sánchez (TCS). While M1 permits the use of three photometric filters $g$, $r$, and $z_\text{s}$ at the same time, M2 observes in four band-passes simultaneously, namely $g$ (400-550 nm), $r$ (550-700 nm), $i$ (700-820 nm), and $z_\text{s}$ (820-920 nm) \citep{Narita_2015, Narita_2019}. M1 and M2 are equipped with 1024$\times$1024 pixel CCDs with a pixel scale of  0.36 and 0.44 arcsec/pixel, resulting in a field of view of 6.1$\times$6.1 arcmin and 7.4$\times$7.4 $\text{arcmin}$, respectively. Both instruments are used for follow-up observations of transiting exoplanets by for example TESS for confirmation and/or validation, and studies of exoplanet atmospheres. The data are reduced and optimized with the M2 transit pipeline \citep{Parviainen_2020}. The pipeline covers the reduction of generic (nontransit) photometry, transit analysis, and more specific TESS follow-up analysis. Moreover, the pipeline tags exposures where any of the pixels inside a photometry aperture for a star are close to the linearity limit of the CCD. The
exposures where either the target star or any of the comparison stars may be saturated are excluded from the analysis. The uncertainty on the data points is inferred using a rolling median with sigma clipping equivalent to the flux error determined by the PDCSAP pipeline for Kepler and TESS data. We estimate the average white noise scatter during the fitting in the pipeline, assuming the uncertainties do not change significantly during the night. 

\begin{figure}[!ht]
    \centering
    \includegraphics[width = \linewidth]{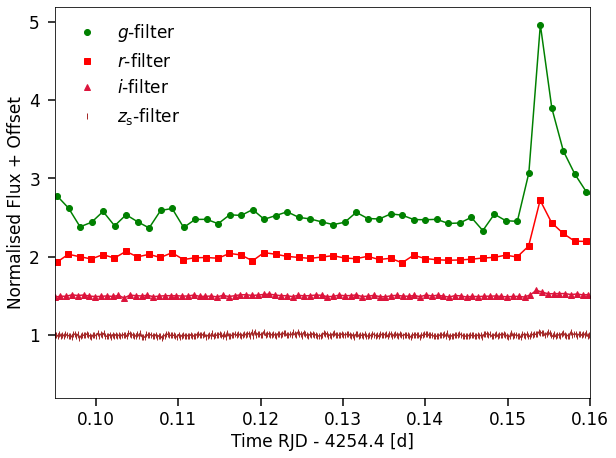}
    \caption{Light curve in all four filters of MuSCAT2 (M2) of Flare 1: The flare is visible in the $g$, $r$, and $i$ filter, whereas the S/N in the $z_\mathrm{s}$-filter is too low. To better distinguish between the four M2 channels, we add constant values to the flux. The time is given in days in reduced Julian date (RJD), RJD = BJD - 2454833 d where BJD is the barycentric Julian date. } 
    \label{fig:observation}
\end{figure}

For the analysis, we used 48 nights of M2 between 2018 and 2020, and 11 nights of M1 between 2016 and 2020, with an observation of 1-3 hours per night for both instruments. Thus, we had a total observation time of four days per filter for the M2 instrument; we refer to Table \ref{tab:obs} for details.

%Twenty out of the 68 nights of M2 could not be used due to dome failure, weather and other instrumental issues, yielding a total number of forty-eight nights of observation. 

\begin{figure}[!ht]
    \centering
    \includegraphics[width = \linewidth]{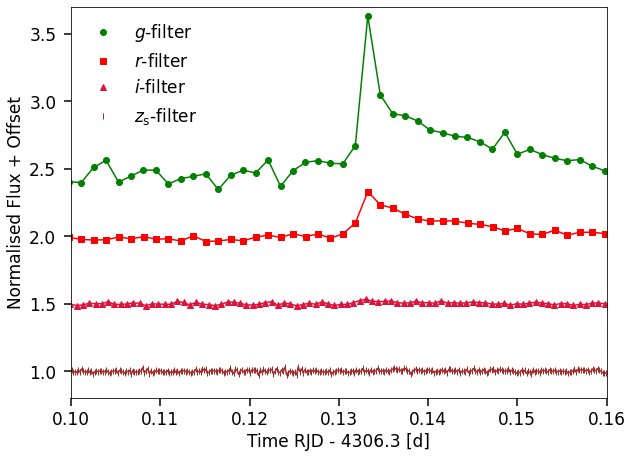}
    \caption{Light curve in all four filters of MuSCAT2 (M2) of Flare 2: The flare is visible in the $g$- and $r$-filter, whereas the S/N in the $i$- and $z_\mathrm{s}$-filter are too low. Time and flux are as indicated in Figure \ref{fig:observation}} 
    \label{fig:observation2}
\end{figure}

We also consider the K2 data set of TRAPPIST-1 \citep{Howell_2014}. \cite{Paudel_2018} analyzed the short-cadence data of the K2 Campaign 12 \citep{Gilliland_2010} and report 39 flares. An overview of the data of TRAPPIST-1 for M1, M2 and K2 is given in the Appendix in Table \ref{tab:obs}. In Figure \ref{fig:observation} and Figure \ref{fig:observation2} both flare light curves of M2 are shown in all four photometric bands.

\section{Methods}
\label{sec:Methods}
\subsection{Flare detection}
\label{sec:flaredection} 

For the flare detection we relied on the \texttt{AltaiPony} package \citep{Ilin_2021b}. The selection criteria were adopted from \citet{2015_Chang}, where every outlier with three subsequent data points exceeding the local  $\sigma$-level by a factor of three, was considered a flare candidate. $\sigma$ represents the local standard deviation. To discriminate the real flares among the candidates found by the algorithm, the standard method is a flare injection recovery (hereafter FLIR), which constrains the probability of a given candidate being recovered in a given light curve. However, as the M2 data for TRAPPIST-1 suffer from strong variation in the photometric scatter due to weather, moon and telescope issues, we would have to perform the FLIR on every single observation night independently. For this reason, we first detected the flares with \texttt{AltaiPony} and afterwards, instead of performing FLIR on all light curves, we visually selected the flare candidates and adopted FLIR for M2 example light curves for each pass-band. The results of the FLIR of the M2 example light curves were used to explore the lower detection limit in bolometric energy. For M2 we obtained a detection limit of $E_\mathrm{bol} = 0.74 \times 10^{30}$ erg, see Appendix 
\ref{sec:flaredec_lim}.

\begin{figure}[!ht]
    \centering
    \includegraphics[width = \linewidth]{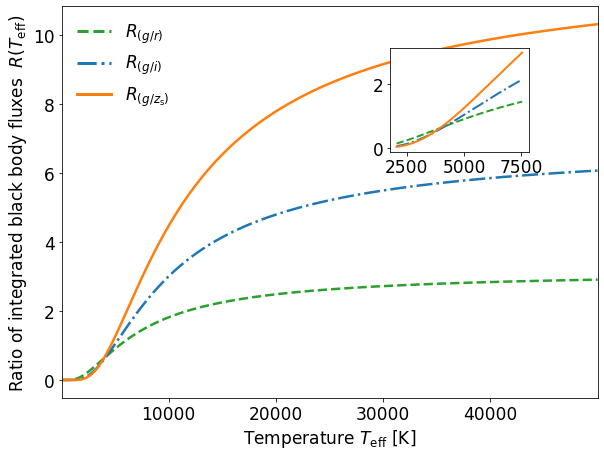}
    \caption{Ratio R of integrated fluxes observed in the MuSCAT band-passes uniquely determines the effective black body temperature of a flare. A black body of temperature $T_\mathrm{eff}$ is separately multiplied by the spectral response functions of the MuSCAT filters to produce the ratios $R_j$. We obtain different sensitivities for each filter pair:  $R_{(g,r_\mathrm{s})}$ is theoretically more sensitive in temperatures below $5000$ K while $R_{(g,z_\mathrm{s})}$ may be sensitive to temperatures up to $50000$ K.}  
    \label{fig:R_tempfunction}
\end{figure}

\subsection{Flare temperatures}
\label{subsec:Flare_Temperatures} 

The color temperature of a stellar flare is defined as the effective continuum temperature associated with a black body inferred from its spectral properties \citep[e.g.][and references therein]{Howard_2020}. To estimate the color temperatures for MuSCAT flares we applied two different methods. 
First, the MuSCAT instruments provide the opportunity with their different photometric filters to measure the spectral energy distribution (SED) of flare events, that is the energy emitted by the flare per unit area, time and wavelength. We attempted to extrapolate the SED leaving the effective temperature and the flaring area as free parameters. The flaring area parameter $a$ is given as a fraction of stellar radius and is the part of the stellar surface where the continuum emission of the flare is emitted. We assume it to be circular. The applied black body model can be found in the Appendix equation (\ref{eq:bbmodel}). To avoid confusion, we refer to the effective temperatures estimated from the SED as SED temperatures $T_\mathrm{SED}$. Second, we followed the methodology from \citet{Howard_2020} c.f. Section 5, to infer the flare color temperature using only two different filters. The global flare temperature $T_\mathrm{glob}$ is defined as the black body effective temperature linked to the total amount of flux of the flare, whereas the peak temperature $T_\mathrm{peak}$ is the temperature associated with the average temperature inside the FWHM of the flare \citep{Howard_2020}. The temperatures can be estimated for each multi-color data point yielding a time evolution in flare temperatures. 

As we did not have any simultaneous flare observed with K2 at different bands, this analysis was done only for the MuSCAT instruments.

\paragraph{SED temperature method:}
%\begin{itemize}
%    \item[1.] We compute the total observed flux from the flare in every MuSCAT passband. We infer the flux from the best-fit flare profile, cf. Section \ref{subsec:ED_calc}, and the filter-specific quiescent flux of the star. We assume that the quiescent flux is constant over time. 
%    \item[2.] To account for the distance of the star and make it comparable to the black body temperatures associated with the surface, we applied flux scaling by the areas related to the distance and radius of the star. We
 %   include the filter sensitivity by dividing the filter-specific flux by the total filter throughput, i.e. they are under the spectral response function. 
%    \item[3.] The MuSCAT SED for any given flare is then fitted with a black body with the effective temperature as a free parameter. 
%\end{itemize}

\begin{itemize}
    \item[1.] We computed the flux of TRAPPIST-1 in each MuSCAT filter using the corresponding segments of the mega-MUSCLES semi-empirical model \citep{Wilson_2021} and calibrated the data to absolute fluxes using the Gaia DR2 catalog \citep{2018_Evans}. For the calibration we followed \citet{2021_Ilin_c}, (c.f. Section 2.4.3). First, we integrated the mega-MUSCLES spectra over the Gaia G-band response function, and normalized the SED to the Gaia G-band flux for TRAPPIST-1. Integrating over the MuSCAT-band response curves, we obtained the fluxes in the MuSCAT2-bands. 
    \item[2.] We then fit the relative flare light curve with the flare template \texttt{aflare1} \citep{Davenport_2014}. From the best-fit flare profile (cf. Section \ref{subsec:ED_calc}), we calculated the percentage flux change of TRAPPIST-1 during the flare. 
    \item[3.] The flare SED was then obtained by multiplying the percentage flux change of the flare with the quiescent flux for any MuSCAT filter. 
    \item[4.] We fit the flare SED with a black body model leaving the effective temperature and the flaring area as free parameters, c.f. Appendix Section \ref{subsec:model_BB}.
\end{itemize}

\par This task can be performed for both the total flare flux, that is, the flux arising from the combined rise, peak and decay phase, and the peak flare flux, which is the flux emitted at the peak of the flare.

\paragraph{Two-filter temperature method from \cite{Howard_2020}:}
\begin{itemize}
    \item[1.] We computed the radiation spectrum of a black body with temperature $T_\mathrm{eff}$ as a function of wavelength $\lambda$. We multiplied all filters separately with the black body spectrum and integrated them over the M1/M2 wavelength range to obtain the passband-specific flux. To account for the filter sensitivity we normalized the fluxes by dividing each calculated passband-specific flux by the total filter throughput. 
    \item[2.] We took the ratios $R_j$ between the flux observed in all different filters, namely M2 $g$, $r$, $i$ and $z_\mathrm{s}$ where  \newline $j \in [(g/r), (g/i), (g/z_\mathrm{s})]$. By convention, we took every single ratio such that the flux of the filter with a higher central wavelength was in the denominator and the flux of the filter with a lower central wavelength is in the numerator. In total we had three ratio functions to probe the color temperatures $[R_{(g/r)},R_{(g/i)}, R_{(g/z_\mathrm{s})}]$. 
    \item[3.] This process was repeated for effective black body temperatures  $T_\mathrm{eff} = [50,50000]$ K with $5$ K steps. 
    \item[4.] To compute the effective black body temperatures of the observed flares, we inverted the different ratio functions such that we had the temperature as a function of the respective filter ratio $T_\mathrm{eff}(R_j)$. 
\end{itemize}

In Figure, \ref{fig:R_tempfunction} the ratio functions $R_j$ are plotted for all different filter combinations for the MuSCAT filters as a function of effective black body temperature $T_\mathrm{eff}$. The $R_j({T_\mathrm{eff}})$ functions indicate that the retrievable color temperature information is limited. Where $R_j({ T_\mathrm{eff}})$ is getting asymptotic, we reach the $R_j$ specific sensitivity limit; see Table \ref{tab:asymlim}. Beyond these limits, we cannot give reasonable estimates of the color temperature. 
In comparison, the ratio functions approach the asymptote faster if the total covered wavelength range between the two filters is smaller. %For flares with high temperature this results in a more difficult temperature estimation towards the red.  %\hl{The reason for these asymptotes is that the MuSCAT filters are distributed in the back end of the Planck curve for a 9000-10000 $^\circ K$ flare, which is usually adopted. This is why we observe a smaller change in ratios for higher color temperatures, which already implies a more difficult color temperature estimation. , this is confusing and not helping} 

Following, \cite{Hawley_2003,Howard_2020,2020_castKleint} we calculated the wavelength-specific flux $F_{\lambda}$, 

\begin{align}
    F_{\lambda} = \frac{A_\mathrm{flare}B_{\lambda}(T_{\mathrm{flare}})}{d^2}
\end{align}

where $A_\mathrm{flare}$ is the area of the flare, $B_{\lambda}(T_{\mathrm{flare}})$ is the black body spectrum of the flare with color temperature $T_\mathrm{flare}$, and $d$ is the distance of the star. $A_\mathrm{flare}$ does not depend on $\lambda$ \citep{Gunther_2020, Howard_2020}, taking the ratios between two passband fluxes yields: 

\begin{align}
\centering
    \frac{F_{\lambda_1}}{F_{\lambda_2}} \approx \frac{ B_{\lambda_{1}}(T_\mathrm{flare})}{ B_{\lambda_{2}}(T_\mathrm{flare}) } = R_{j} 
    \label{eq:fraction}
\end{align}

With equation (\ref{eq:fraction}) we used the ratio functions to infer the temperatures for the observed flares. Following \cite{Howard_2020}, this could be done in three different ways. We first investigated the flare temperature epoch by epoch, and then the global flare temperature. Finally, and importantly, we examined the flare peak temperature. We employ the calculation of all three measures and probe their consistency with the SED temperatures from the total flux and the peak flux.

\subsection{Equivalent duration}
\label{subsec:ED_calc}

\begin{figure}[!ht]
    \centering
    \includegraphics[width = \linewidth]{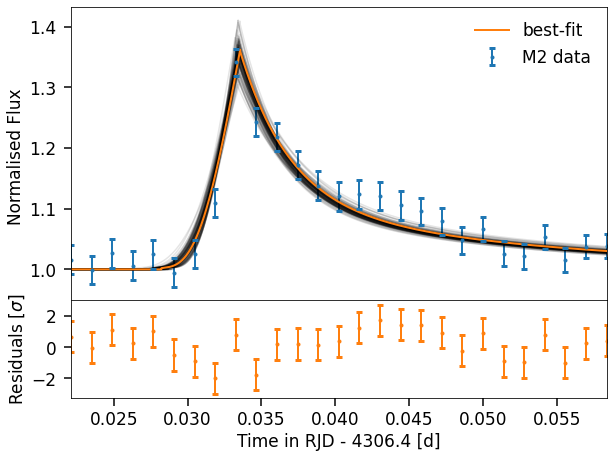}
    \caption{Flux profile fit of Flare 2 with M2 $r$-filter data: The orange curve indicates the best-fit using the flare template \texttt{aflare1}. The black lines show 100 random samples from the Markov chain. The lower panel shows the residuals between the observation and the model. The corresponding posterior probability distribution of the MCMC-fit is shown in Figure \ref{fig:corner_fl2} of the Appendix. The lower panel reveals systematic errors in the residuals which might be due to superimposed flares or oscillations. This is further discussed in Section \ref{sec:Discussion}. } 
    \label{fig:ResidualFit}
\end{figure}

The equivalent duration ($ED$) is defined as the amount of time it would take the star in its quiescent state to produce the same amount of energy that is released in a flare event \citep{Gershberg_1972, Hunt-Walker_2012}. Thus, it is the time integral of the dimensionless flux, 
\begin{align}
     ED = \int {\frac{F_\text{flare} - F_\mathrm{q}}{F_\text{q}}dt}  ,
\end{align}

where $F_\mathrm{q}$ is the quiescent flux of the star and $F_\mathrm{flare}$ is the flux observed during the flare.
\noindent To estimate the equivalent duration, we fit the flare template \texttt{aflare1} \citep{Davenport_2014}
to each selected flare and thus acquire a best-fit approximation using an MCMC approach \citep{Mackey_2013}. The analytic expressions for the flare template are given in Section \ref{subsec:Analytical_template} in the Appendix. We used 5000 steps in the MCMC chain with 40 walkers, and discard the first 1000 steps as the burn-in phase. To put constraints on the parameters for the flare template (amplitude, peak time and full width at half maximum (FWHM)) we define Gaussian prior probability distributions, see appendix \ref{Gauss_p}. 
After the fitting process, we integrate the flare template with the best-fit parameters for each flare event over time and thus obtain the \textit{ED}, making use of the trapezoidal sum for integration. In Figure \ref{fig:ResidualFit} a flare profile fit is shown for the $r$-band. \par
Stellar flares are often multi-peaked \citep{Gunther_2020} or show quasi-periodic pulsations (QPP) \citep{2003_Mathioudakis,2005_Mitra}. We focused in this work on modeling classical single-peaked flare events and disregard periodic oscillations in the decay phase in our analysis because both would not affect our temperature estimation; c.f. Section \ref{subsec:oscillation}.

 %The model includes both the rising phase of a flare and the decay phase, 

%\begin{align}
%    F_{rise} = 1 + 1.941(\pm 0.008)t_{1/2} − 0.175(\pm 0.032)t^2_{1/2} −  \\ \notag 2.246(\pm 0.039)t^3_{1/2} − 1.125(\pm0.016)t^4_{1/2}  \\ 
%    F_{decay} = 0.6890(\pm0.0008) \exp{−1.600(±0.003) t_{1/2}} \\ \notag + 0.3030(\pm0.0009) \exp{−0.2783(\pm0.0007) t_{1/2}}
%\end{align}

%where $t_{1/2}$ 

\section{Results}
\label{sec:Results}

We found no flares in the M1 data and two flares were observed in all four filters in the M2 data. Flare 1 had a bolometric energy of $E_\mathrm{bol} = 2.99 \times 10^{31}$ erg and occurred on August $25$ 2020 and Flare 2 had a bolometric energy of $E_\mathrm{bol} = 1.51 \times 10^{31}$ erg and occurred on October $15$ 2020. Both flares are shown in all passbands in Figure \ref{fig:observation} and Figure \ref{fig:observation2}, respectively.

\subsection{SED temperatures}
\label{subsec:SED_Temp}

To infer the SED for our M2 flares, we used the methods described in Section \ref{subsec:Flare_Temperatures}. 
\par
\begin{figure}[!ht]
    \centering
    \includegraphics[width = \linewidth]{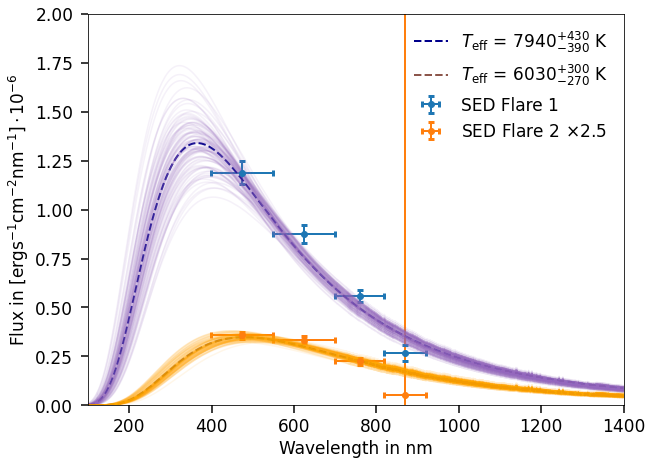}
    \caption{SED for the two observed flares in M2. The SEDs are fit with a model where the temperature $T_\mathrm{eff}$ and the area parameter $a$ are left as free parameters using  \texttt{emcee} again. The flux in $z_\mathrm{s}$-filter is slightly inconsistent with the best-fit black bodies for both flares. The uncertainties of the model are indicated with shaded areas. The uncertainty bars in wavelength should be understood as the respective filter widths. The SED for Flare 2 is multiplied by a constant factor of 2.5 for a better visual comparison between the two SEDs. The purple and orange lines indicate 100 randomly drawn samples from the Markov chain and mark the uncertainty of the fitted dashed best-fit curves. } 
    \label{fig:SED}
\end{figure}

\begin{figure}[!ht]
    \centering
    \includegraphics[width = \linewidth]{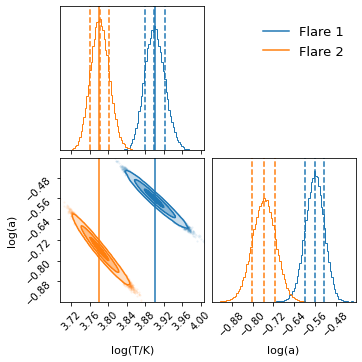}
    \caption{Posterior probability distribution for both flare samples: The sample of Flare 1 is represented in blue and the samples of Flare 2 are in orange. The sampling is done in logarithmic parameter space such that both the temperature $T$ and the area parameter $a$ are given here in logarithmic scale.  }
    \label{fig:SED_corner}
\end{figure}

 
The inferred SEDs for both flares are shown in Figure \ref{fig:SED}. The uncertainties were propagated from the best-fit of the flare flux profiles and the uncertainties of the mega-MUSCLES spectrum, as well as the uncertainties on the Gaia flux. To find a black body temperature consistent with the empirically measured filter fluxes, we used our black body model (c.f. equation (\ref{eq:bbmodel}) of the Appendix) and employed a fit to the data using \texttt{emcee} \citep{Mackey_2013}. We used 40 walkers and 25000 steps in the chain and discarded the first 5000 steps as the burn-in phase. The convergence was checked visually by plotting the chains against step number and by computing the integrated autocorrelation time  \citep{Sokal_1996,2010_Goodman,Mackey_2013}. We report autocorrelation times of roughly 35 steps for each trial. The used prior probability distributions can be found in \ref{tab:Prior}. In Figure \ref{fig:SED_corner} the combined posterior probability distributions for both flare samples are shown and reveal a satisfactory resolution of the expected degeneracy between the area parameter and the flare temperature. The posterior distributions overlap marginally for the flare area parameter but indicate that the flare temperatures are indeed different between the two flares. The best-fit temperatures are $T_\mathrm{SED} = 7940 \substack{+430 \\ -390} $ K  for Flare 1 and  $T_\mathrm{SED} = 6030 \substack{+300 \\ -270} $ K for Flare 2. The same task is performed for the peak temperatures yielding  $T_\mathrm{SEDp} =  13650 \substack{+1550 \\ -1250}$ K  for Flare 1 and $T_\mathrm{SEDp} = 8300 \substack{+700 \\ -550}$ for Flare 2. 
The best-fit area parameters to the SEDs are $a = 0.266 \substack{+0.023\\-0.021}$ for Flare 1 and $a = 0.171 \substack{+0.019\\-0.017}$ for Flare 2. And for the peak SED $a_\mathrm{p} = 0.536 \substack{+0.062\\-0.070}$ for Flare 1 and $a_\mathrm{p} = 0.213 \substack{+0.027\\0.031}$ for Flare 2. The coverage of the stellar surface is represented in Table \ref{tab:areas} and indicates that the flares cover up to $28\%$ of the surface of TRAPPIST-1 at the peak and in the combined rise, peak and gradual phases up to $8\%$.

%(0.5363145349867854, 0.06144288590413621, 0.06939288299392166)

%(0.21327007585458763, 0.026513075088009097, 0.030277020470174443)

%\begin{table}
%      \caption{Best-fitted SED temperature from black body fit to the spectral energy distribution of the two MuSCAT flares.} 
%         \label{tab:SE}
%     $$ 
 %        \begin{array}{p{0.25\linewidth}l p{0.25\linewidth}l}
  %          \hline
   %         \noalign{\smallskip}
    %        & T_{eff} [ K]  \\
     %       \hline
     %       \noalign{\smallskip}
     %       \text{flare 1} & 5590 \substack{+130 \\ -110}  \\
     %       \text{flare 2} & 5000 \substack{+120 \\ -90}  \\
     %       \hline
    %        \noalign{\smallskip}
%         \end{array}
%     $$ 
%\end{table}

%125.83131039]), array([104.06371626
%array([112.91484268]), array([83.15549882]))

\subsection{Flare temperature by epoch}
\label{subsection:Flare_epoch}

\begin{figure}[!ht]
    \centering
    \includegraphics[width=\linewidth]{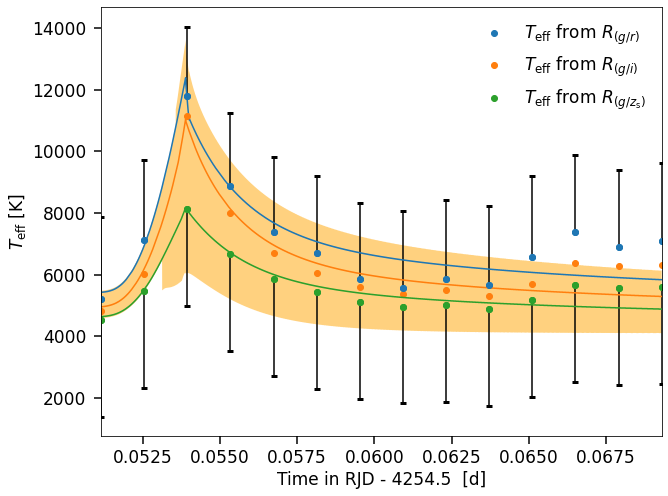}
    \caption{Temperature evolution of Flare 1: Using the ratio functions $R_j$, we infer for each time step a flare color temperature by comparing the observed flux ratios with the theoretical black body flux ratios in between two MuSCAT filters using the $\chi^2$ method. Different ratio functions $R_j$ yield different temperature evolutions. The uncertainty range is indicated by black markers and is valid for all three temperature evolutions. The lines represent the smoothly varying model for the flare color temperature which is derived from the best-fit parameters of the MCMC fitting to the flare light curve using the flare template \texttt{aflare1}. The orange shaded area gives the overlaid uncertainty of all three models.}
    %and also the uncertainties vary, decreasing towards "redder" ratios.
    \label{fig:temp_evol}
\end{figure}

The flare temperature was calculated epoch-by-epoch to acquire the time evolution of the color temperature, see Figure \ref{fig:temp_evol}. Thus, fluxes were computed for every time step and compared with each other as presented in Section \ref{subsec:Flare_Temperatures}.

Because the observations in different passbands yield different cadences, we had to bin the observed flux to compare the theoretical black body fluxes between two filters. The uncertainty in each time step was propagated from the uncertainty of the filter-specific fluxes to the ratio functions. \par 
To obtain a smoothly varying model for our epoch-by-epoch color temperature, we made use of our best-fit flare flux profile model. We fit the flare template \texttt{aflare1} to the flare flux profile for each flare and MuSCAT passband, obtaining a smooth flux model for the flare event. Subsequently, we performed the same steps as before to retrieve the color temperatures from these flux profiles, cf. Section \ref{subsec:Flare_Temperatures}. We used our color temperature model to compute the global flare temperature, cf. Section \ref{subsubsection:global_flare_temper}. \par

%\begin{figure}
%    \centering
%    \includegraphics[scale = 0.25]{figures/flare_temperatures_epoch_error.png}
%    \caption{Temperature evolution of a flare using $R_{(g,i)}$ ratio function: The figure shows the same flare as before but only uses one ratio function to illustrate the uncertainties on the flare temperatures in each time-step. The errors are typically in the order of $10^3$ K for all ratios $R_i$.}
%    \label{fig:flares_epocherr}
%\end{figure}

\subsection{Global vs peak flare temperature}
\label{subsubsection:global_flare_temper}

If signal-to-noise ratio (S/N) is low, the temperature evolution is not a precise measure for the color temperature of the flare and other measures should be used \citep{2020_Namekata}. A more sophisticated approach to determining the temperature of a given flare makes use of the total instead of the epoch-by-epoch flux. Using our best-fit approximation for the flare flux profile (see Section \ref{subsec:ED_calc}), we calculated the total flux in each filter by integrating the flare template \texttt{aflare1}  with our best-fit parameters obtained for each flare. We compared the acquired flux ratios for each filter pair and applied the ratio functions $R_j$ to yield an estimate of the ratio-specific global flare temperature. The same can be done for the peak fluxes that is, the flux within the FWHM of the flare. Table \ref{tab:global_obs} shows the global and the peak flare temperatures, respectively, for every possible filter combination in comparison to the derived SED temperatures. Also, we combined our results for all filters by weighting the temperatures by their uncertainties.

%We measure the peak temperature, defined as the
%mean temperature during the flare FWHM, \cite{Howard_2020}. As they showed for the 2 minutes cadence of TESS and \cite{1980_Mochnacki} showed for 10 seconds cadence, for low cadence the peak temperature coincides with the maximum temperature of the temperature time evolution. This is why for M2 with a typical cadence below 2 minutes, we adopt the observational peak temperature to be the maximum of the temperature evolution. Another way to give a better estimate of the peak temperature is to use the model flare temperature, see section \ref{subsubsection:Flare_epoch}. We can infer the peak temperature by averaging the temperatures from the model in the FWHM of the flare time and thus yield a unique peak temperature for every filter combination from the model.

\begin{table}[!ht]
\caption{Flaring area parameter and respective flaring areas for both flares shown in fractional units of the stellar radius. }    
 \label{tab:areas} % title of Table
\centering                                      % used for centering table
\begin{tabular}{c c c }          % centered columns (4 columns)
\hline                       % inserts double horizontal lines
\noalign{\smallskip}
Global & \hspace{25mm} Fraction of $R_\mathrm{star}$ \\
            & $\text{a} \times R_\mathrm{star}$ & $\text{A} \times \pi R_\mathrm{star}^2$ \\
\hline
\noalign{\smallskip}
Flare 1 & $0.266 \substack{+0.023\\-0.021}$ & $0.071 \substack{+0.001 \\ -0.001 }$ \\
Flare 2 & $0.171 \substack{+0.019\\-0.017}$ & $0.029 \substack{+0.001 \\ -0.001}$ \\ 
\noalign{\smallskip}
\hline 
\noalign{\smallskip}
Peak & \hspace{25mm} Fraction of $R_\mathrm{star}$ \\
            & $\text{a} \times R_\mathrm{star}$ & $\text{A} \times \pi R_\mathrm{star}^2$ \\
\hline
\noalign{\smallskip}
Flare 1 & $0.536 \substack{+0.062\\-0.070}$ & $0.288 \substack{+0.004 \\ -0.005 }$ \\
Flare 2 & $0.213 \substack{+0.027\\-0.031}$ & $0.045 \substack{+0.001 \\ -0.001}$ \\ 
\noalign{\smallskip}
\hline 
\noalign{\smallskip}                                            %inserts single line
\end{tabular}
\end{table}

\begin{comment}

\begin{table}[!ht]
      \caption{Flaring area parameter and respective flaring areas for both flares shown in fractional units of the stellar radius. } 
         \label{tab:areas} 
     $$ 
         \begin{array}{p{0.25\linewidth}l p{0.25\linewidth}l}
            \hline
             \noalign{\smallskip}
            \text{Global} & \hspace{12 mm} \text{Fraction of}  \hspace{2mm} R_\mathrm{star}    \\ 
            \noalign{\smallskip}
            & \text{a} \times R_\mathrm{star} & \text{A} \times \pi R_\mathrm{star}^2 \\
            \noalign{\smallskip}
            \hline
            \noalign{\smallskip}
            Flare 1 & 0.266 \substack{+0.023\\-0.021} & 0.071 \substack{+0.001 \\ -0.001 }  \\ 
            Flare 2 & 0.171 \substack{+0.019\\-0.017 } & 0.029 \substack{+0.001 \\-0.001 } \\ 
            \noalign{\smallskip} 
            \hline
            \noalign{\smallskip}
            \text{Peak} & \hspace{12 mm} \text{Fraction of} \hspace{2mm} R_\mathrm{star}  \\ 
            \noalign{\smallskip}
            & \text{a} \times R_\mathrm{star} & \text{A} \times \pi R_\mathrm{star}^2 \\
            \noalign{\smallskip}
            \hline
            \noalign{\smallskip}
            Flare 1 & 0.536 \substack{+0.062\\-0.070} &  0.288 \substack{+0.004\\-0.005} \\
            \noalign{\smallskip}
            Flare 2 & 0.213 \substack{+0.027\\0.031} &  0.045 \substack{+0.001\\-0.001} \\
            \noalign{\smallskip}
            \hline
         \end{array}
     $$ 
   \end{table}
\end{comment}

\begin{table}[!ht]
  \caption{Global and peak flare temperatures retrieved from the observation of the two TRAPPIST-1 flares. } 
         \label{tab:global_obs}
         \centering 
         \begin{tabular}{c c c}
            \hline
            \noalign{\smallskip}
                 & \hspace{15mm} \text{Global $T_\mathrm{glob}$ [K]}   \\
            \text{Ratio}  & \text{Flare 1} & \text{Flare 2} \\
            \noalign{\smallskip}
            \hline
            \noalign{\smallskip}
            $R_{(g,r)}$ &  $7000 \pm 2500$  & $6200 \pm 2100$ \\ 
            $R_{(g,i)}$ & $6300 \pm 3100$  & $5700 \pm 2600$ \\ 
            $R_{(g,z_\mathrm{s})}$ &  $5600 \pm 3300$ & $5100 \pm 2800$ \\  
            \noalign{\smallskip} 
            \text{Weighted Sum} &   $6450 \pm 650$   &  $5800 \pm 500$     \\
            \hline
            \noalign{\smallskip}
            \text{From SED} & $7940 \substack{+430 \\ -390}$  & $6030 \substack{+300 \\ -270}$\\
            \noalign{\smallskip}
            \hline
            \noalign{\smallskip}
            \hline
            \noalign{\smallskip}
                 & \hspace{15mm} \text{Peak $T_\mathrm{peak}$ [K]}   \\
            \text{Ratio}  & \text{Flare 1} & \text{Flare 2} \\
            \noalign{\smallskip}
            \hline
            \noalign{\smallskip}
            $R_{(g,r)}$ &  $11800 \pm 2300$  & $8420 \pm 1880$ \\ 
            $R_{(g,i)}$ & $11200 \pm 3100$  & $7350 \pm 2600$ \\ 
            $R_{(g,z_\mathrm{s})}$ &  $8200 \pm 3250$ & $6300 \pm 2600$\\  
            \noalign{\smallskip}
            \text{Weighted Sum} &   $10800  \pm  1550$  &  $7650 \pm 900$ \\
            \hline
            \noalign{\smallskip}
            \text{From SED} & $13620 \substack{+1520 \\ -1220}$ & $8290 \substack{+660 \\ -550}$ \\ 
            \noalign{\smallskip}
            \hline
            \noalign{\smallskip}
          \end{tabular}
          The global flare temperatures take the total flux into account, whereas the peak flare temperatures are calculated using the flux within the FWHM. For comparison, the temperatures inferred from the SED of the individual flares are also shown.
     
\end{table}

\section{Discussion}
\label{sec:Discussion}

Flare temperatures on active M-dwarf stars (dM3e-dM4.5e) were previously derived from spectroscopic and photometric observations \citep{2013_Kowalski, 2021_Johnson} - $T_\mathrm{glob} = 9000-10000$ K and $T_\mathrm{peak} = 9000-14000$ K. It is usually assumed that flares on ultra-cool dwarfs have similar temperatures \citep[e.g.][and references therein]{2013_Kowalski, Paudel_2018, Gunther_2020, Glazier_2020}. In this work, we showed with real photometric data that this assumption cannot be transferred to TRAPPIST-1, an ultra-cool M-dwarf with spectral type M7.5V. Thus raises the question of whether or not flares on late M-dwarfs should in general be modeled with cooler temperatures than early M-dwarfs. 

\begin{comment}
\begin{table}[!ht]
      \caption{Usually assumed temperatures for late M-dwarfs flares. The global temperature refers to the temperatures associated with the total amount of flare flux and thus includes the raise, peak and gradual phases and the peak temperature only takes the peak flux into account. In comparison, our SED temperatures were shown. \hl{Tabelle in text} } 
         \label{tab:Temper_bf} 
     $$ 
         \begin{array}{p{0.25\linewidth}l p{0.25\linewidth}l}
            \hline
            \noalign{\smallskip}
                  &  \text{Temperatures in} \hspace{2mm} [K]     \\
            \noalign{\smallskip}
            \hline
            \noalign{\smallskip}
            \text{T}_\mathrm{glob} & 9000 - 10000   \\ 
            \text{T}_\mathrm{peak} & 9000 - 14000 \\ 
            \noalign{\smallskip} 
            \hline
            \noalign{\smallskip}
            \text{T}_\mathrm{SED} & 5770 - 8380          \\
            \text{T}_\mathrm{SEDp} & 7750 - 15200 \\  
            \noalign{\smallskip}
            \hline
         \end{array}
     $$ 
   \end{table}
\end{comment}

\subsection{SED versus two-filter method}
\label{subsec:cons_methods}

%Nevertheless, the expectation from previous studies was that TRAPPIST-1 should show a flare of 10000 \pm 1000 K (Kowalski) and 11000 K \p 3500 \m 2600, Howard, such that our statement that we observed lower than expected flare temperatures remains valid. 

The results from both the SED and the two-filter method imply that the global and peak temperature for the observed TRAPPIST-1 flares are lower than expected from the empirical flare temperatures used in modeling the energy budget of a flare \citep[e.g.][and references therein]{2010_segura,2013_Kowalski,2019_Tilley, Howard_2020}, except for the peak temperature of Flare 1, where the temperature from both applied methods is consistent with the literature. The two methods are consistent within a $2-\sigma-$confidence interval; see Table \ref{tab:global_obs} for Details. 
The color temperature evolutions are shown in Figure \ref{fig:temp_evol} and Figure \ref{fig:temp_evolutio}. For both flares, the different ratio functions are consistent within the uncertainties.  
Comparing both methods, the clear advantage of the SED method is the simultaneous use of all filter fluxes and the simultaneous fit of the area parameter. As a consequence, the SED method is not affected by the disadvantage that we do not have a blue filter in our data set. 
The uncertainties on the two-filter methods could be improved by MCMC sampling of the theoretical black body flux values rather than using a simple $\chi^2$ reduction. However, this is beyond the scope of this paper as the two-filter method was only meant to prove the temperature consistency of our SED method. \par
\noindent For Flare 1, the peak temperature for both applied methods is between 9000 and 14000 K and is therefore consistent with the literature. For Flare 2, the peak temperature lies within 2-$\sigma$ from the literature value for both methods. The reason for these consistent peak temperatures could be twofold. The fitting of the temperatures becomes more difficult toward higher temperatures because of the lack of a bluer photometric filter. This is partly monitored by the larger uncertainties that we obtain for the peak SED temperatures in comparison to the global SED temperatures. On the other hand, if the flare has high peak SED temperatures but lower global SED temperatures, this could imply that the cooling mechanisms are different in the atmosphere TRAPPIST-1 in comparison to earlier M-dwarfs. 

%The SED method yields a more accurate flare temperature and thus is recommended for future flare studies with the MuSCAT instruments. Nevertheless, in similar future studies, we strongly encourage the usage of blue filters, e.g. the U-band, if possible. 

\subsection{SED temperature uncertainties}

For the SED method, the estimated uncertainties were inferred by Gaussian uncertainty propagation taking the uncertainty from the flare flux profile and the uncertainty on the mega-MUSCLES spectrum. The marginalized uncertainties on the temperature were estimated directly by taking the $84^\mathrm{th}$ and 16$^\mathrm{th}$ percentiles of the MCMC samples. Observing Figure \ref{fig:SED} closely, the uncertainties on the $z_\mathrm{s}$-flux are the largest, extending over the whole SED flux regime for Flare 2. The $z_\mathrm{s}$-flux is more sensitive to the flux-profile-fitting component, because of the low flare S/N in the $z_\mathrm{s}$-band. For Flare 2 the low S/N prevented an accurate fit to the flare flux profile in the  $z_\mathrm{s}$-band and therefore we obtain large uncertainties. \par
Strong emission lines can also contribute to uncertainties. For instance, the peak flare flux could be contaminated by H$\alpha$  and H$\beta$ emission by up to 10$\%$ and 8.8$\%$ of the total flux \citep{2013_Kowalski}, respectively. Taking the maximum of the H$\alpha$ contribution into account, the $r$-filter flux was reduced for the global calculation and for the peak calculation by roughly $30\%$ and $26\%$, respectively. Indeed, this correction for the H$\alpha$ contribution results in higher flare temperatures because the $r$-filter flux is corrected toward lower flux values, meaning that the SED is more likely to be fit with higher temperatures, unless another line contamination is  considered. This could be compensated by the H$\beta$ emission in the $g$-filter. We reduced the flux in $g$-filter by approximatly $50\%$ and $20\%$ for the total and peak calculations, respectively. As a consequence of the flux reduction in the $g$-filter the overall shape of the SED is shallower for both cases such that it is more likely to fit with lower temperatures than that of the reported black body. Contamination from both lines is probably is present in our data and therefore they will cancel each other out to a certain degree. However, it is impossible to correct the SED for the real H$\alpha$ and H$\beta$ contributions. By omitting line emission in our calculations our uncertainties on the inferred black body temperatures are increased, but also it does not indicate a clear bias since the strongest line emission are prone to cancel each other out to a certain degree in the fitting procedure

\begin{comment}

Fitting our data without the $r$-band or without $g$-band, respectively,  the results for both peak and global temperatures are consistent with the values given in table \ref{tab:global_obs}. Therefore, it is likely that we overestimated the H$\alpha$ contribution in the previous calculation where we only took H$\alpha$ contribution into account.
\end{comment}
\par

\subsection{Flare location}

The position of the flare on the stellar surface has an impact on the received flux and therefore might alter the SED. Because of the rotation of TRAPPIST-1 with period $P \in [1.4,3.295]$ days  \citep{2016_Gillom,2017_Vida, 2017_Luger}, the flare might move over the stellar surface during its emission. The observed flares have durations of less than 30 minutes and therefore their angular movement on the surface is $\Delta \alpha \in [4.5,1.9] ^\circ$. The projected angles are even smaller, such that the time was not sufficient for the observed flares to move from the limb to the center or vice versa. Thus, we assumed that the projected surface area of TRAPPIST-1 is constant. Applying simple laws of limb darkening \citep{Claret_2000}, it was evident that the flux at the limbs reduced already at a line-of-sight angle of roughly 60 degrees to half its value at the center. Because the M2 data are given in relative units we used the mega-MUSCLES Spectra for TRAPPIST-1 to derive the SED of the flare. Bluer colors are more affected by limb darkening; see Figure (\ref{fig:Center_2}) of the Appendix. As discussed in Section \ref{subsec:BB_model}, the flare temperature determines the slope of the SED and therefore the reduction of the flux close to the limb can be neglected. If we assume that the flares occurred close to the limb, the shape of the SED would result in a higher temperature than if they occurred in the vicinity of the center; see Figure \ref{fig:Center_2}. Therefore, we assumed that our best-fit temperatures are consistent with lower temperatures and independent of the exact position on the stellar surface.

\subsection{Black body model}
\label{subsec:BB_model}

Our model relies on two main characterizing parameters, the flare temperature $T_\mathrm{eff}$ and the flare area parameter $a$; see Section \ref{subsec:model_BB} in the Appendix. The black body distribution determines the shape and slope of the model. Here, $a$ is the constant of proportionality between the observed flux and our model.  
Using the MuSCAT data directly, the flare area parameter would not be a measure for the real covered radius fraction, because the MuSCAT data are given in relative units. We therefore calibrated our observation to the Gaia absolute fluxes. Qualitatively, the flare area parameter shows the expected behavior. Flare 1 emits more flux and therefore has also a larger flare area parameter than Flare 2. Because the flare temperature is the parameter that determines the shape of the SED, our model can still determine $T_\mathrm{eff}$ satisfactorily even if no absolute calibration is done but only the relative MuSCAT calibration. We tested this behavior for both flares obtaining the same temperatures but different flaring area parameters. Also, we performed a model sensitivity test for Flare 1 by multiplying the flare flux by arbitrary constant factors; see Figure \ref{fig:model_sen}. Multiplying the SED with constant factors resulted in a change in the flare area parameter but left the temperatures unchanged.

\subsection{Residuals in flare profiles}
\label{subsec:oscillation}

In Figure \ref{fig:ResidualFit} the fit shows residual systematic errors. This could be due to complex flare events with multiple peaks or QPPs. Any superimposed flare would increase flux. As discussed in the previous section, the amount of flux is regulated by the flare area parameter and therefore a superimposed flare would not contaminate our flare temperatures, unless it changes the shape of the SED. Additionally, we employed a sinusoidal fit to the residuals with period $P = 29.02 \pm 6.53
$ min. Typical durations for QPPs range from several seconds to tens of minutes for M-dwarf flares \citep[e.g][]{2016_Van, 2021_Million}. Because our period is consistent with the typical QPP periods we considered our systematic errors to be explained by a QPP. The flux oscillations do not imply the flare temperatures and therefore the main result of our paper remains valid. For this reason, we neglected the observed QPP for our calculations.

\subsection{Flare temperatures per spectral type} 
\label{subsec:Flaretemp_spectype}

\begin{figure}[!ht]
    \centering
    \includegraphics[width = \linewidth]{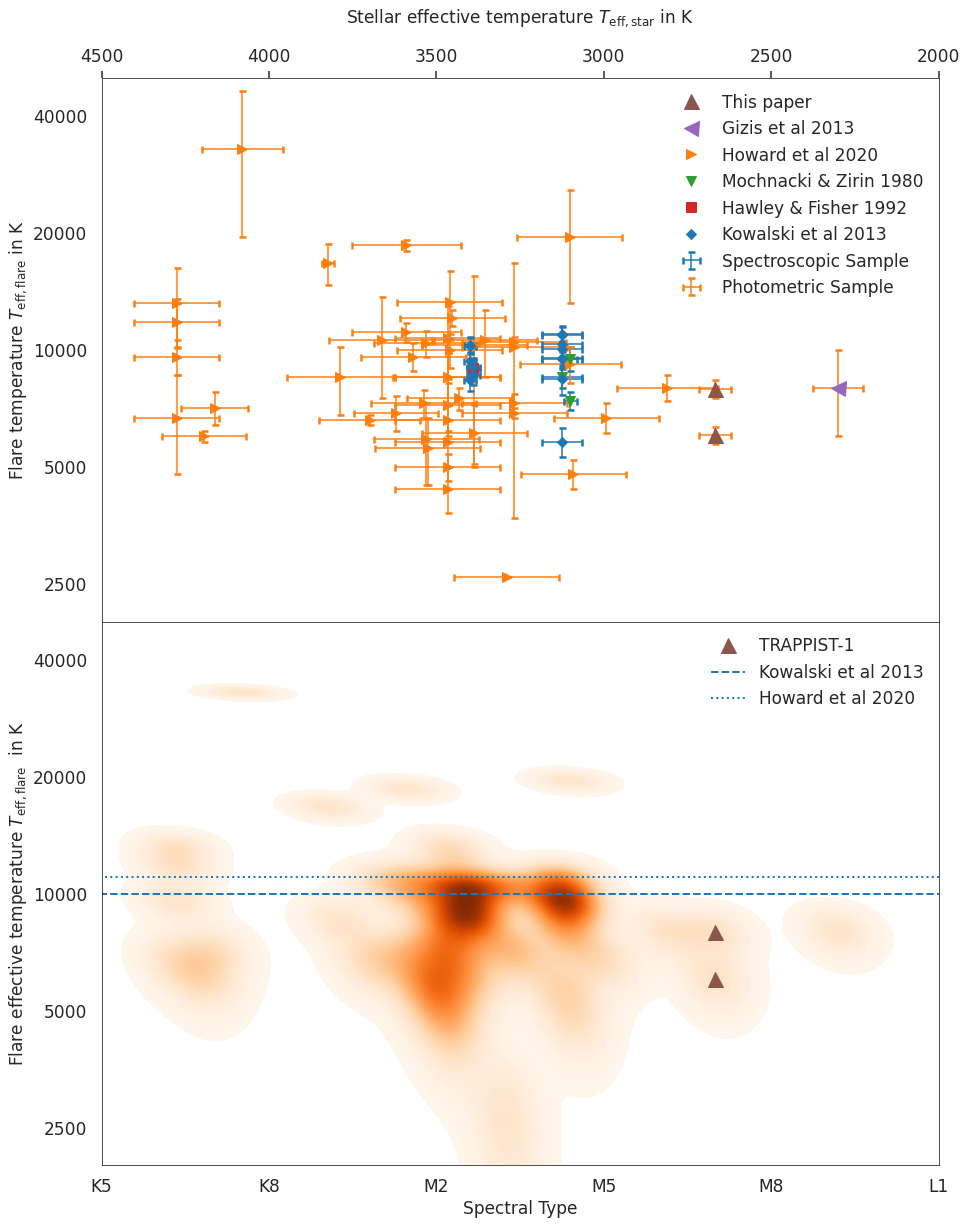}
    \caption{Spectral type and stellar effective temperature versus total flare temperature for different photometric and spectroscopic studies: The upper panel uses the stellar effective temperature if available from the TESS input catalogue \citep{Stassun_2019}. For the Kowalski sample, we use conservative uncertainties of $\pm 500$K. The lower panel shows a density map of flare temperatures per spectral type that is consistent with the upper stellar effective temperatures. The flare temperature axis is represented in log-scale. The two TRAPPIST-1 flares are overplotted and the blue horizontal lines represent the average temperatures from the two biggest temperature samples.}
    \label{fig:fig9}
\end{figure}

\begin{figure}[!ht]
    \centering
    \includegraphics[width=\linewidth]{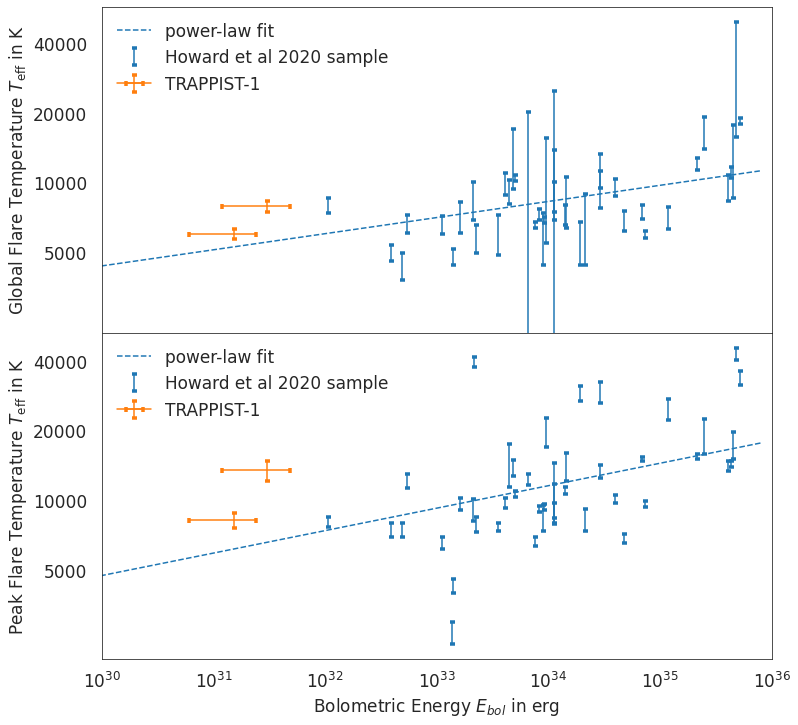}
    \caption{Energy-temperature relation in log-log representation: In the upper panel the global flare temperatures are plotted versus the bolometric energy, whereas in the lower panel the peak temperatures are used. The flares from \cite{Howard_2020} are indicated in blue and our two TRAPPIST-1 flares are shown in orange. The blue dashed line indicates a power-law fit to the data.}
    \label{fig:fig11}
\end{figure}

Figure \ref{fig:fig9} shows the flare continuum temperatures for TRAPPIST-1 and other M- and L-dwarfs from different photometric and spectroscopic studies \citep{1980_Mochnacki,1991_Hawley,2013_Kowalski, Gizis_2013,Howard_2020}. The upper panel reveals a possible trend for low-mass stars. The hottest flares seem to get cooler toward lower masses, such that the higher right corner is not populated by flares. This could be explained by a selection bias. As \cite{Howard_2020} pointed out in their energy-temperature relation, more energetic flares seem to be hotter; see Figure \ref{fig:fig11}. These higher energetic flares are more likely for earlier type M-dwarfs  \citep{2019_Howard,Gunther_2020} and therefore hotter flare temperatures $T_\mathrm{eff}>10000$ K could be caught more frequently for early M-dwarfs than for later types like TRAPPIST-1. \par 
Toward the tail of the low-mass-stars, there are only three flare temperatures measured. All three of them have bolometric flare energies between $10^{31}$ and $10^{32}$ and are in the lower energy tail of the energy-temperature relation. Therefore, it is difficult to verify the trend with spectral type, as there are not yet enough measurements of ultra-cool dwarf flares. Later type stars turn fully convective at roughly spectral type M4V and have cooler atmospheres, i.e. surface temperatures of below $T_\mathrm{eff} = 3200$ K \citep{chabrier_1997,reid_2013}. If the trend with spectral type is physical it might be due to a different flare generation process for fully convective stars in comparison to stars with radiative cores. Also, the heating mechanisms of the lower chromosphere driven by the accelerated particles could be different in late M-dwarfs.  Without further observation and modeling of ultra-cool M-dwarf chromospheres, this discussion remains speculative. \par
Figure \ref{fig:fig11} shows the energy-temperature relation including the two TRAPPIST-1 flares for bolometric energies.  We fit a power law of the form $\log_{10}(T_\mathrm{eff}) = \alpha \cdot E_{bol} + \beta$ with two coefficients $\alpha$ and $\beta$ to the \cite{Howard_2020} sample, including the two TRAPPIST-1 flares, and obtain for the total temperatures $\alpha_{tot} = (6.98 \pm 0.03) \cdot 10^{-2} , \beta_{tot} = 1.5 \pm 0.4$ and for the peak temperatures  $\alpha_{peak} = (9.67 \pm 6)\cdot 10^{-2} , \beta_{peak} = 0.8 \pm 0.7$ . Both peak and global temperatures of the TRAPPIST-1 flares are consistent with the uncertainty of the relation. 
However, we argue that the correlation is complete only if the flaring area is included in the relation because we expect that the heating process is affected not
only by the energy released from the magnetic field but also by the size of the area that is heated. This could explain the large observed scatter especially for the peak temperatures. To obtain a correlation between energy-temperature and area, we propose to measure area and temperature simultaneously and dependent on each other in future similar studies; for example, by taking the dominant cooling mechanism in the chromosphere we could model how the temperature is distributed at the flare footpoints and how the area evolves then with time in dependence of the temperature. However, this is beyond the scope of the present work.

From the density map in the lower panel of Figure \ref{fig:fig9}, it follows that cool flares $T_\mathrm{eff} < 8000$ K are also present in early spectral types and are not exclusively occurring on late M-dwarfs. Nevertheless, it appears that they are less frequent than flare temperatures around $T_\mathrm{eff} = 10000$ K. Following the energy-temperature relation, surprisingly, lower flare temperatures are not denser on the map, as they should occur more frequently. This could be explained by selection effects due to  observational limitation for example.

\begin{figure}[!ht]
    \centering
    \includegraphics[width=\linewidth]{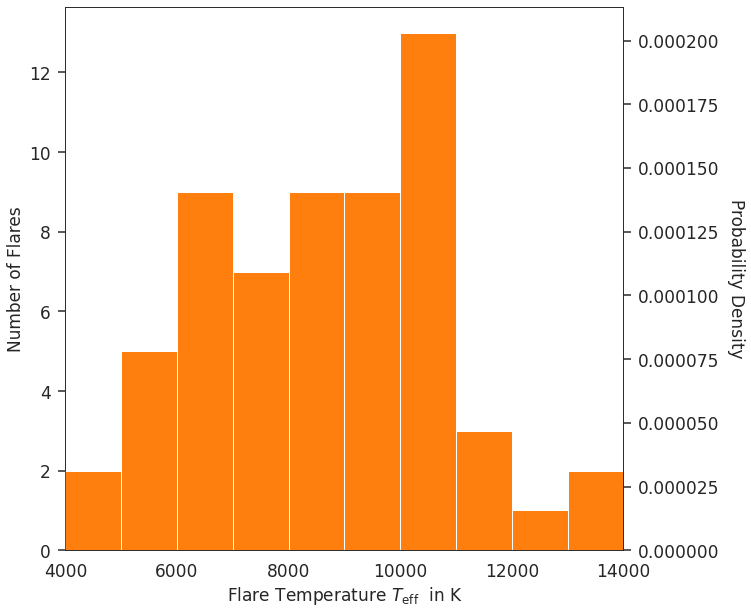}
    \caption{Histogram of total flare temperatures from the spectroscopic and photometric observations shown in Figure \ref{fig:fig9} with the peak of the distribution close to 10000 K. }
    \label{fig:fig10}
\end{figure}

Figure \ref{fig:fig10} a histogram of measured total temperatures from the same sample as in Figure \ref{fig:fig9}. The distribution reveals the peak at 10000 K as expected. Our TRAPPIST-1 flares with 6030 and 7940 K, respectively, seem less likely but not at all unlikely. The histogram shows that the numbers of flares in a given temperature range increase, until 11000 K is reached. Afterwards, there appears to be a cut. Between 11000 and 12000 K the observed numbers decrease strongly. It appears that flares with $T_\mathrm{eff} = 10000-11000$ K are the most likely observed ones and flares with temperatures below 10000 K appear to be more likely than flares with $T_\mathrm{eff} > 11000$ K.  This could be explained by the empirical energy-temperature relation mentioned above. Flares with higher energy are less likely to occur and therefore also flares with higher temperatures are less frequent. 

Taking the energy-temperature relation and the flare frequency distribution (FFD; see following Section), we predict that we would have to observe TRAPPIST-1 for at least 15 days to find a flare with temperatures above 10000 K. Taking this into account, we conclude that the flares observed on TRAPPIST-1 show lower temperatures than expected when compared to the average temperatures from previous studies \citep{Hawley_2003,2013_Shibayama,2013_Kowalski,Howard_2020}.  However, if the flare frequency distribution and the energy-temperature relation are considered it is not surprising that we observed lower flare temperatures for TRAPPIST-1. 

\subsection{Flare frequency distribution and habitability}
\label{subsec:Flare_freq}

One goal of our analysis was to update the FFD of TRAPPIST-1 \citep{Paudel_2018, Glazier_2020} and show the implications of a different flare temperature. The FFD is the cumulative rate of flares per time unit
\citep{Gershberg_1972} and is frequently described by a power law \citep{Davenport_2016, Paudel_2018, Gunther_2020, Glazier_2020}. In log-log space, this power law has a linear appearance,  
\begin{align}
    \log_{10}{\nu} = \alpha \log_{10}{E_\mathrm{bol}} + \beta ,
    \label{eq:flare_fr}
\end{align}

where $\nu$ is the occurrence rate of the flares in a given time unit, $E_\mathrm{bol}$ is the bolometric energy $\alpha$ indicates the slope and $\beta$ is the y-intercept in log-log space. The bolometric energies were calculated as presented in Section \ref{subsec:Flare_energies} of the Appendix and we adopted a value of 7000 $ \pm 500$ K for the flare temperature, which is consistent with the results of this paper. \par 
First, we took the flares found by \cite{Paudel_2018} in the K2 light curve and computed the bolometric energies by adopting their $ED$s following Section \ref{subsec:Flare_energies}.

%from the Kepler passband energies by adapting the inverse of the energy fraction $1/c = 3.1$ and multiplying it with the Kepler passband specific energies. For details on the fraction see appendix \ref{subsec:Filter_E}. 

\begin{figure}[!ht]
    \centering
    \includegraphics[width=\linewidth]{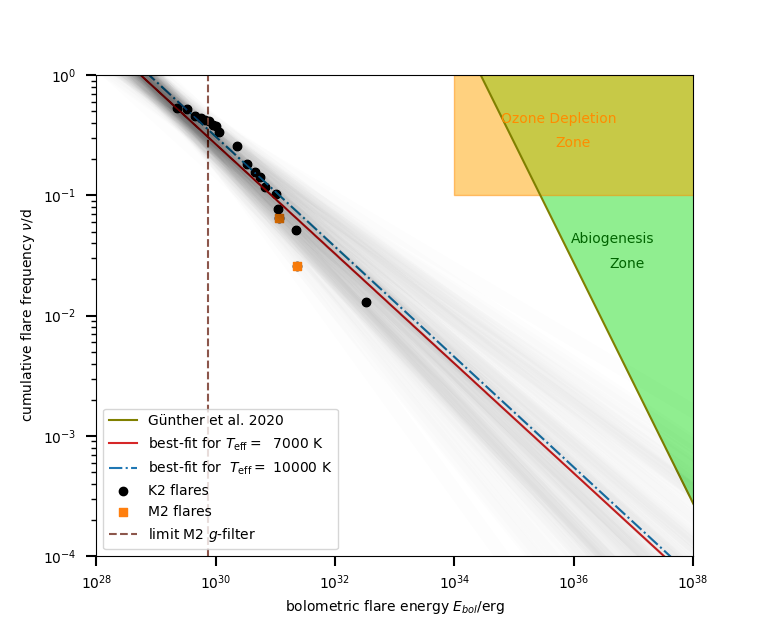}
    \caption{Flare frequency distribution for K2 and M2 data. The FFD for TRAPPIST-1 is updated by our flares indicated in orange. The black points are adopted from \citep{Paudel_2018}. The best fit to the power law is chosen from 2500 MC trials, here only 100 random samples are shown to mark the uncertainty of the model. Also overplotted are the abiogenesis zone in green \citep{Rimmer_2018,Gunther_2020} and the ozone-depletion zone in yellow \citep{2019_Tilley, Gunther_2020} for TRAPPIST-1.}
    \label{fig:FFD}
\end{figure}

The updated FFD for the combined information of M1, M2 and K2 is shown in Figure \ref{fig:FFD}. We observe a truncated power law with a turnoff at roughly $10^{30} \mathrm{erg}$. This is a result of flares with lower energies close to the K2 detection limit as discussed for Kepler by \citet{2014_Hawley}. The theoretical detection limit for M2, shown by the vertical dashed line, is given by the $g$-band; see Section \ref{sec:flaredection}. The orange squares indicate the two M2 flares, while the black points represent the K2 flares \citep{Paudel_2018}.

Using the power law ansatz for the flare frequency equation (\ref{eq:flare_fr}), we fit the flare frequency with the MCMC approach of \citet{Wheatland_2005} implemented in \texttt{AltaiPony}. The best-fit to the data is indicated by the red solid line, as well as 100 samples of the 2500 MC trials. By extrapolating the FFD to larger bolometric energies, we explored the FFD into the regime of low flare occurrence, that is higher energies. We analyzed potential danger to the atmospheres of the TRAPPIST-1-companions and possible abiogenesis. The blue dash-dotted line indicates the best-fit if the bolometric energies were calculated with a black body temperature of 10000 K. Comparing the two best-fit solutions with the different adopted temperatures, we obtained $\alpha_{7000 \mathrm{K}}  1.45 \pm 0.20$ , $\alpha_{10000 \mathrm{K}} = 1.46 \pm 0.24$. The higher temperature led to a slightly steeper slope. However, the impact is not significant because both slopes are consistent with each other.
In the abiogenesis zone of an MS star, specific prebiotic chemical reactions are possible that enable the synthesis of ribonucleic acid (RNA) \citep{Rimmer_2018, Gunther_2020}. The flare frequency $\nu$ needed to trigger and sustain prebiotic chemistry is given by \citet{Gunther_2020}; %Günther paper 
\begin{align}
 \nu \geq 25.5 \hspace{0.1 cm}\mathrm{days}^{-1} \left (\frac{10^{34}\mathrm{erg}}{{E_\mathrm{U}}}  \right) \left( \frac{R_\mathrm{star}}{R_\odot}\right)^2 \left(\frac{T_\mathrm{star}}{T_\odot}  \right)^4
\end{align}

where $E_\mathrm{U}$ is the U-band energy, $R_\mathrm{star}$ the stellar radius and $T_\mathrm{star}$ the stellar surface temperature. In Figure \ref{fig:FFD}, the abiogenesis zone is plotted in green, adopting the surface temperature and radius of TRAPPIST-1 of $T = 2648 \pm 26 $ K \citep{Wilson_2021} and $R = 0.1192 \pm 0.0013$   $R_\odot$ \citep{agol_2021}. Both the red and blue lines do not intersect with the abiogenesis zone, indicating, that the high energetic flux from TRAPPIST-1 flares is likely to be insufficient to trigger and sustain prebiotic chemistry. \cite{Glazier_2020} came to the same conclusion. \par 

%We used the Sloan u'-band filter with central wavelength $\lambda_c = 359.6  nm$ and width $\Delta \lambda=  570 nm$ \citep{Bessel_2005} and calculated the fraction of the energy in the U-band by using equation (\ref{eq:const}) to obtain the conversion constant $c = 0.058$ where \cite{Gunther_2020} found $c = 0.076$, due to a different photometric system. 

%Since the bolometric energy is temperature dependent (see Equation (\ref{eq:flarear})), the necessary flare frequency $\nu \propto 1/T_{eff}^4$ must decrease with increasing flare color-temperature. This means that to trigger abiogenesis we would need either less high energetic or many low energetic flares to hit the atmospheres of the exoplanets. 

The yellow shaded area in Figure \ref{fig:FFD} marks the ozone depletion zone around TRAPPIST-1, where frequent high-energy flares may erode exoplanet atmospheres and lead to subsequent surface sterilization. \cite{2019_Tilley} modeled the impact of M-dwarf flares on the atmosphere of an Earth analog planet. These authors argued that Earth's ozone layer should erode if flares with bolometric energies $\geq 10^{34}$ erg hit the atmosphere at a frequency of $\nu_{34} \geq 0.4 \hspace{1mm} \text{day}^{-1}$. \cite{Gunther_2020} gave a lower limit for the frequency $\nu_{34} \geq 0.1 \hspace{1mm} \text{day}^{-1}$. We adopted the more conservative approach of \citet{Gunther_2020}, which also allowed us to remain consistent with the analysis of \citet{Glazier_2020}. As the best-fit did not intersect with the ozone-depletion zone, we conclude that the UV flux is insufficient to deplete a complete Earth-like ozone layer. \par 

%\footnotetext{\url{https://old.aip.de/en/research/facilities/stella/instruments/data/johnson-ubvri-filter-curves} on Aug 14 at 16:00 p.m.}

%As previous studies have shown, TRAPPIST-1's activity does not give rise to the loss of ozone for its habitable zone planets nor is it enough to trigger and drive prebiotic chemistry, see \cite{Glazier_2020} from whom the style of this Figure was adopted.

\section{Conclusion}
\label{sec:Summary}

We used 59 nights of multi-color photometric observations of MuSCAT1 and MuSCAT2 to search for flares on TRAPPIST-1. We found two flares in the MuSCAT2 light curves. Moreover, we discovered that the black body temperatures associated with the total emitted flux are likely to be cooler than previously suggested for ultra-cool M-dwarfs. We inferred temperatures of $T_\mathrm{SED} = 7940 \substack{+430 \\ -390} $ K for Flare 1 and  $T_\mathrm{SED} = 6030 \substack{+300 \\ -270} $ K for Flare 2. We obtained peak temperatures $T_\mathrm{SEDp} =  13620 \substack{+1520 \\ -1220}$ K  for Flare 1 and $T_\mathrm{SEDp} = 8300 \substack{+660 \\ -550}$ for Flare 2. The observed peak black body temperatures for TRAPPIST-1 also show marginally cooler temperatures for Flare 2, while they are consistent with the literature for Flare 1. It could be that different cooling mechanisms in the stellar atmosphere are responsible for this behavior and that cooling is more efficient toward later spectral types. This remains speculative without further spectroscopic measurements and modeling. Lower black body temperatures for flares lead to different UV surface fluxes and therefore have an impact on the habitability estimation on exoplanets around ultra-cool M-dwarfs, even though we showed the bolometric energies remain marginally affected. While using the 9000-10000 K assumption one has to keep in mind, that flares with lower temperatures are more likely to occur and thus, we suggest using a flare frequency temperature distribution to account for the different flare temperatures in the future.
\par 
Furthermore, we conclude that, based on our data, it is not possible to verify whether the lower observed temperatures on TRAPPIST-1 are intrinsic to ultra-cool dwarfs or  are the result of lower energies generated for these late-type stars, making it more likely to observe lower flare temperatures as suggested by the energy-temperature relation. Our prediction of roughly 15 days of observation to observe a flare with $T_\mathrm{eff} > 10000$ K is a considerable task for future work and would also show whether or not the energy-temperature relation can be expanded to the low-mass end of M-dwarfs. 
\par
Our SED method proves to be suitable for estimating flare temperatures and areas. It is a good trade-off between previous multi-color photometric- and spectroscopic approaches because of its efficiency and low cost. Even though spectroscopic methods enable the resolution of the specific line emission, we show that our model allows us to infer precise temperatures and areas associated with stellar flares.  \par 
    %\item[5.] While the results are exciting, they should be used with caution. Due to the lack of a blue filter in our analysis, the Planck curve is not caught completely in our spectral energy distributions. Thus, we cannot be strictly sure that flare temperatures are that cold. 
Further multi-color observations are needed to put constraints on other M-dwarf flares in order to verify whether or not the observed behavior is indeed a general characteristic of ultra-cool M-dwarfs. The reported trend of lower flare temperatures toward later spectral types could be investigated by a thorough observation campaign of several late M-dwarfs and early L-dwarfs to populate the low-mass tail in Figure \ref{fig:fig9}. M-dwarfs with a surface temperature of $\sim 3000$ K are particularly suitable targets as they show enhanced activity in comparison to other M-dwarfs and earlier type stars \citep{Gunther_2020}. We recommend temperature measurements of similar stellar types to TRAPPIST-1 with bluer filters. Also, we encourage the spectroscopic observation of ultra-cool M-dwarf flares in order to infer H$\alpha$ and H$\beta$ contributions, for example, to improve the results.

%\footnotetext{The reason to fit the FFD with a power-law rather than a broken power-law is that the cumulative flare rate close to the detection limit is underestimated \citep{Hawley_2003}. }

\begin{acknowledgements}
    A.J.M. would like to thank the anonymous referees for their thoughtful review of the work.
    \textbf{
    This article is based on observations made with the MuSCAT2 instrument, developed by ABC, at Telescopio Carlos Sánchez operated on the island of Tenerife by the IAC in the Spanish Observatorio del Teide. }
    This article was supported by the Excellence Cluster ORIGINS which is funded by the Deutsche Forschungsgemeinschaft (DFG, German Research Foundation) under Germany's Excellence Strategy - EXC-2094 - 390783311. 
    E.I. acknowledges support from the German National Scholarship Foundation.
    E. E-B. acknowledges financial support from the European Union
    and the State Agency of Investigation of the Spanish Ministry of Science and
    Innovation (MICINN) under the grant PRE2020-093107 of the Pre-Doc Pro-
    gram for the Training of Doctors (FPI-SO) through FSE funds. N.N. acknowledges the support by JSPS KAKENHI Grant Number JP18H05439, JST CREST Grant Number JPMJCR1761. M.M. is supported by Grant-in-Aid for JSPS Fellows, Grant Number JP20J21872. We acknowledge financial support from the Agencia Estatal de Investigación
    of the Ministerio de Ciencia, Innovación y Universidades through
    projects PID2019-109522GB-C53
    The authors want to thank Maximilian Günther and Ward Howard for helpful discussions based on their expertise on stellar flares on M-dwarfs. 
\end{acknowledgements}


%
\bibliographystyle{aa}
\newpage
\bibliography{literature}

\begin{appendix} %First appendix
%\appendix 

\section{Flare detection limits of the MuSCAT 2 instrument}
\label{sec:flaredec_lim}

\begin{table}[!ht]
      \caption{Observation of TRAPPIST-1 for the different  instruments.}
         \label{tab:obs}
         \begin{center}
         \begin{tabular}{c c c}
            \hline
            \noalign{\smallskip}
            \text{Instrument}      &  \text{Total Time [d]} &   \text{Cadence [s]}  \\
            \noalign{\smallskip}
            \hline
            \noalign{\smallskip}
            M1 & (1.2, 0.8, 0.2) & (85, 64, 68) $z_\mathrm{s}$,$r$,$g$ \\ 
            M2 & (4.2, 3.4, 4.5, 4.2) & (39, 60, 63, 61)  $z_\mathrm{s}$,$i$,$r$,$g$  \\ 
            K2 & 75  & 60 \\ 
            \noalign{\smallskip}
            \hline
            \noalign{\smallskip}
         \end{tabular}
          \end{center} 
      Shown are the total observation time and the cadence for each instrument. The cadence is averaged over the whole observation period of the respective instrument. We note that for the MuSCAT instruments the cadence varies strongly in each observation year such that the average should be understood as a rough estimation. We have an observation in 2016 for K2, 2016, 2017 and 2020 for M1 and 2017, 2018, 2019 and 2020 M2. 
\end{table}

\begin{figure}[!ht]
    \centering
    \subfigure[]{\includegraphics[width=0.24\textwidth]{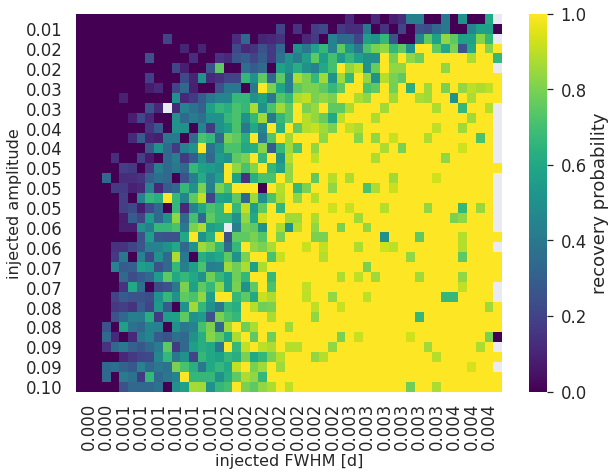}} 
    \subfigure[]{\includegraphics[width=0.24\textwidth]{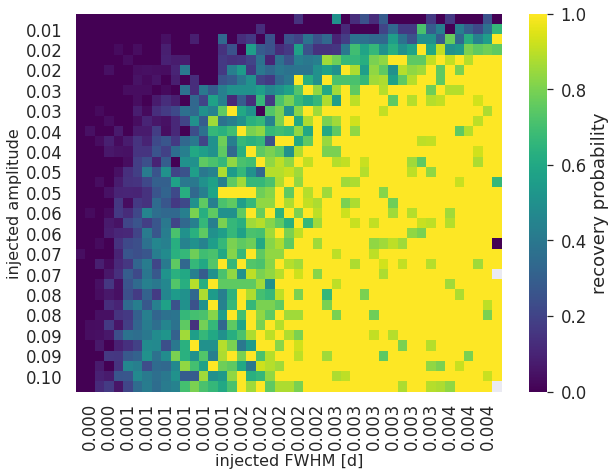}} 
    \subfigure[]{\includegraphics[width=0.24\textwidth]{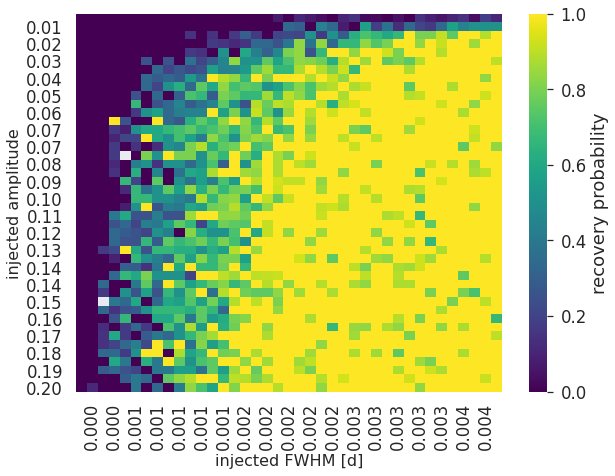}} 
    \subfigure[]{\includegraphics[width=0.24\textwidth]{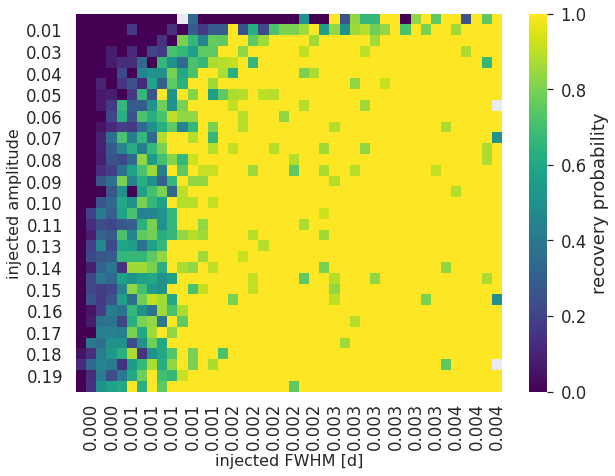}} 
    
    \caption{Flare injection recovery via heat maps for each MuSCAT 2 (M2) filter, Panels (a),(b),(c), and (d) correspond to the M2 $g$, $r$, $i$ and $z_s$ photometric filters. The detection limits in amplitude and FWHM of the flares are for all bands visible and marked by the dark regions where the recovery probability is low. } 
    \label{fig:Detection_limit} 
\end{figure}

To infer the detection limits, we generated a mock light curve, for each M2 filter by adopting normalized flux and adding Gaussian noise with a mean equal to the median of the respective noise level of all M2 observations for TRAPPIST-1. The typical observation time is calculated from Table \ref{tab:obs} by adapting the total observation time per filter and their averaged cadences. The results from FLIR are shown in Figure \ref{fig:Detection_limit}, where the injection of 20000 classical single-peaked events with the \cite{Davenport_2014} template \textit{aflare1} were visualized for each filter. The FLIR was executed iteratively, i.e. only one flare was injected per iteration in the light curve to avoid overlapping of injected flares. In Figure \ref{fig:Detection_limit}, the dark regions mark low recovery probability, while the yellow areas indicate high recovery probability \footnotemark. \par 
We defined the detection limit as the region where the FLIR delivers $100 \%$ recovery probability, for the first time, because there we were certain that any flare event with these properties would be detected. This approach may seem too conservative but as the light curve is artificial, it seems appropriate to remain as conservative as possible. The detection limits in terms of bolometric energy, FWHM, and amplitude of the flare are listed in Table \ref{tab:det_lim}. The ranges for FWHM and amplitude implied that the recovery probability was calculated in FWHM and amplitude bins, such that in the given bin with the border of the given ranges, we have $100 \%$ recovery probability. The bolometric energy was then calculated using the central value of these ranges by adopting the equations presented in Section \ref{subsec:Flare_energies}, where the uncertainty was computed by Gaussian error propagation and is constrained by the width of the FWHM and the amplitude for a given filter.

\begin{table}[!ht]
      \caption{The lower detection limits for MuSCAT. }
         \label{tab:det_lim}
         \begin{center}
         \begin{tabular}{c c c c}
            \hline
            \noalign{\smallskip}
            \text{Filter} & \text{FWHM} [$\mathrm{min}$] & \text{Amplitude} & $E_\mathrm{bol}$ [$10^{30} \mathrm{erg}$] \\
            \hline
            \noalign{\smallskip}
            $g$ &  $[5.4 , 5.76]$ & $[0.059, 0.075]$ & $0.74 \pm 0.05$ \\
            $r$ & $[4.32, 4.68]$ & $[0.03, 0.04]$ & $3.12 \pm 0.18$ \\ 
            $i$ & $[3.96, 4.32]$ &  $[0.035, 0.053]$ & $8.13 \pm 0.45$  \\ 
            $z_\mathrm{s}$ & $[2.02,2.16]$ & $[0.04, 0.05]$ & $8.48 \pm 0.81$ \\
            \noalign{\smallskip}
            \hline
            \noalign{\smallskip}
         \end{tabular}
         \end{center} 
         
      The lower detection limits in FWHM and amplitude for all filters associated with the bolometric energy adopting a blackbody temperature of 7000 K. Even though the adopted photometric scatter is lower in $z_s$ and $i$, the detection limits in $g$ and $r$ are lower in terms of energy, because the flare flux from a 7000 K is peaked closer to the blue than the red light.
\end{table}

\footnotetext{https://altaipony.readthedocs.io/en/latest/index.htm on Aug 13th at 13:00 a.m.}

We also had a limit for the detection of long-duration and hence high-energy flares. If the FWHM of the flare exceeds the duration of the observation (1.5-2 hours), the flare was not completely resolvable and the fitting process might not converge as described in Section \ref{subsec:ED_calc}. However, this limit was difficult to quantify because of its multidimensionality. It depends not only on the FWHM of the flare but also on its peak time and was therefore not quantified in this work. Moreover, the detection of such highly energetic flares was unlikely in our data set. By looking at the FFD of TRAPPIST-1 we identify that flares with bolometric energy of $E_\mathrm{bol} > 10^{34}$ occur only once in 100 days. As our total observation time is roughly 5 days with the combined MuSCAT observation we neglected the detection limit of long-duration flares.

\begin{comment}
The detection limits are explored by injecting 20000 classical artificial flares iteratively into a M2 mock light curve, where the flare template of \cite{Davenport_2014} is used. The mock light curves are generated by taking the normalized flux and adding noise from a Gaussian distribution with a mean equal to the median noise level of all M2 observations available for TRAPPIST-1 and the respective filter. The typical observation duration is calculated from Table \ref{tab:obs} by taking the total observation time and cadence for each filter into account. The $z_s$-filter revealed the best time resolution, such that we should observe here the lowest detection limit in FWHM in comparison to the others. The expectation is met if we compare the transition of yellow to blue in all of the four filters. Since we have the lowest photometric scatter in the $z_s$- and $i$-bands in principle, low amplitude flares should also be better visible there. However, the flux we receive for a given flare is higher toward the blue light and will thus lead to higher signal-to-noise ratios in $g$ and $r$, as shown in Table \ref{tab:SNR}, such that the possible recoverable energy is lower towards the blue.
\end{comment}

\section{Flare energetics} 

\subsection{Analytic flare template}
\label{subsec:Analytical_template}

\cite{Davenport_2014} used for the generation of their flare template \texttt{aflare1} flares that occurred on GJ1243 (M4.0Ve). The duration of used flares for the generation of the template was in the range of 20-75 minutes. Therefore, both of our observed flares with durations of over $t > 30$ min fit well in the selected flare range. The analytic formula for the flare flux profiles relies on a combination of polynomial and exponential fits. The rise phase, i.e., $-1t_{1/2} \leq t \leq 0$, is described by a polynomial function \citep{Davenport_2014}:

\begin{align}
F_\mathrm{rise} = 1 + 1.941(\pm 0.008)t_{1/2} - 0.175(\pm 0.032)t^2_{
1/2} \\ \notag  - 2.246(\pm 0.039)t^3_{1/2} - 1.125(\pm 0.016)t^4_{1/2}
\end{align}

with $t_{1/2} = x - t_\mathrm{peak}/FWHM$ being the full width in time at the maximum of the flare and $x$ the observational time \citep{2013_Kowalski}. The decay phase, i.e., $t > 0$, is described by \citep{Davenport_2014}:

\begin{align}
    F_\mathrm{decay} = 0.6890(\pm 0.0008) e^{-1.600(\pm 0.003) t_{1/2}} + \\ \notag 0.3030(\pm 0.0009) e^{-0.2783(\pm 0.0007) t_{1/2}}
\end{align}

Therefore the full template reads: 

\begin{align}
    F_\mathrm{flare} = (F_\mathrm{rise} + F_\mathrm{decay}) \cdot |ampl|
\end{align}

where $ampl$ is the amplitude in the flux of the flare.

\subsection{Gaussian prior probability}
\label{Gauss_p}

For the MCMC fitting of the flare flux profiles, we needed to define prior probability distributions for the three parameters of the flare template \texttt{aflare1}. The estimate of the amplitude $am_\mathrm{rec}$ from  \texttt{AltaiPony} was taken as the mean of the respective Gaussian prior and $0.05$ in terms of normalized flux as a conservative standard deviation. The peak time $t_\mathrm{peak}$ was calculated as the \texttt{argmax} of the normalized flux for a given flare event. We therefore chose the standard deviation of the peak time to be one data point in each observation. The $FWHM_\mathrm{est}$ was only approximated very roughly by dividing the duration of the flare by the arbitrary value of four. For this reason the standard deviation for the FWHM was chosen to be $1/24$ days and the prior was therefore rather uninformative. For an overview of the adapted Gaussian priors see Table \ref{tab:Prior}.  

\begin{table}
      \caption{Gaussian prior for the MCMC-approach for the MuSCAT instruments. } 
        \label{tab:Prior} 
        \begin{center}
        \begin{tabular}{c c}
            \hline
            \noalign{\smallskip}
            \text{Parameter}      &   \text{M1 and M2}   \\
            \noalign{\smallskip}
            \hline
            \noalign{\smallskip}
            $t_\mathrm{peak}$ & $\mathcal{N}(t_\mathrm{peak},\frac{0.2}{60 \cdot 24} \text{d})$  \\ 
            $am_\mathrm{flare}$ & $\mathcal{N}$($am_\mathrm{rec},0.05$) \\ 
            $FWHM_\mathrm{flare}$ &  $\mathcal{N}$($FWHM_\mathrm{est}$,$\frac{1}{24} \text{d})$ \\ 
            \noalign{\smallskip}
            \hline
            \noalign{\smallskip}
            $T_\mathrm{eff}$ & $\mathcal{U}(log(T_\mathrm{eff}),3.3,4.4 \text{K})$ \\
            \noalign{\smallskip}
            $a$ & $\mathcal{U}(log(a),-10,0)$ \\
            \hline
            \noalign{\smallskip}
         \end{tabular}
         \end{center} 
    
    $am_\mathrm{rec}$ refers here to the recorded amplitude given by \texttt{Altaipony} and is given in terms of normalized flux. We use all filters with the same standard deviations. For the MCMC-fits of the SEDs, we use uniform priors for the effective temperature  $T_\mathrm{eff}$ and the fraction of flaring area $a$ in logarithmic space.
\end{table}
   
 \begin{figure}
     \centering
     \includegraphics[width =\linewidth]{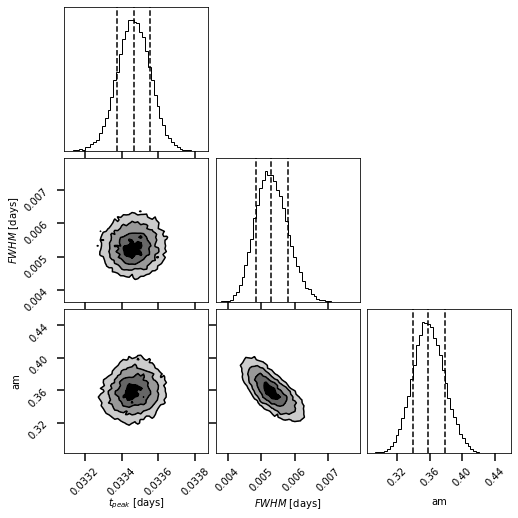}
     \caption{Corner plot of MCMC  flare profile fit of Flare 2 in the $r$ passband. Shown are the correlations between the three parameters of \texttt{aflare1} - peak time, full width at half maximum and amplitude. The dashed lines indicate $15^{\text{th}}$, $50^{\text{th}}$ $84^{\text{th}}$ quantiles. }
     \label{fig:corner_fl2}
 \end{figure}

\subsection{Bolometric flare energies}
\label{subsec:Flare_energies}

In this chapter, we present how the flare energies were calculated. We followed the bolometric flare energy calculation of \citep{2013_Shibayama, Gunther_2020}, where the stellar luminosity and the best-fit flare flux profile are used. 

First, we calculated the bandpass independent area $A_\mathrm{flare}(t)$ of the flare assuming the flare luminosity can be modeled as a black body with constant effective temperature of $T_\mathrm{flare} = 9000 \pm 500 $ K ,  

\begin{align}
    A_\mathrm{flare}(t) = \frac{F_\mathrm{flare}(t) - F_\mathrm{q}}{F_\mathrm{q}} \pi R_\mathrm{star}^2 \frac{\int R_\lambda B_\lambda(T_\mathrm{eff})d\lambda }{\int R_\lambda B_\lambda(T_\mathrm{flare})d\lambda} 
    \label{eq:flarear}
\end{align}

where $R_\lambda$ denotes the instrument response function, $B_\lambda$ the black body radiation, $T_\mathrm{eff}$ the effective temperature of the star in its quiescent state, and $ R_\mathrm{star}$ refers to the stellar radius. The effective flare temperature $T_\mathrm{flare}$ was chosen as a lower conservative limit in order to remain consistent with previous studies \citep{ Davenport_2016, Paudel_2018,Gunther_2020} and references therein, but it was not a correct assumption, because the flare temperature not only changes with time but also depends on the flare phase. Also, we show in this work that the temperature associated with the whole flare event is lower than frequently adopted for ultra-cool M-dwarfs. \par
Having $A_\mathrm{flare}(t)$, we can compute the bolometric flare energy $E_\mathrm{flare}$ of the flare, using its corresponding luminosity,  

\begin{align}
    E_\mathrm{flare} = \int L_\mathrm{flare}(t) dt = \sigma_\mathrm{sb} T_\mathrm{flare}^4 \int A_\mathrm{flare}(t) dt 
    \label{eq:E_flare}
\end{align}

We calculate the uncertainties of the calculation using Gauss uncertainty propagation. Integrations of this section were performed via the trapezoidal sum rule. 

\subsection{Filter-specific flare energies}
\label{subsec:Filter_E}

We calculated not only the bolometric but also the filter-specific energy, in order to infer the detection limits. We obtained the filter-specific energies by adopting a conversion constant from the bolometric energies. Assuming the flare has a bolometric luminosity $L_\mathrm{flare}(t)$ and filter specific luminosity $L_{\lambda,\mathrm{flare}}(t)$, then:

\begin{align}
    L_\mathrm{flare}(t) = \sigma_\mathrm{sb} T_\mathrm{flare}^4 A_\mathrm{flare}(t) \\
    L_{\lambda,\mathrm{flare}}(t) = L_\mathrm{flare}(t) \cdot c
\end{align}

where the flaring area $A_\mathrm{flare}(t)$ is not passband-specific, $T_\mathrm{flare}$ is the effective blackbody temperature from the flare and $c$ denotes the dimensionless conversion constant. By using \mbox{$L = F\pi d^2$}, we obtain: 

\begin{align}
    c = \frac{L_{\lambda,\mathrm{flare}}}{L_\mathrm{flare}} = \frac{F_{\lambda,\mathrm{flare}}}{F_\mathrm{flare}}
    \label{eq:const}
\end{align}

with $F_{flare}$ denoting the flux from the flare event. Because the bolometric energy associated with the flares was defined as in equation (\ref{eq:E_flare}), we derived the passband-specific energy using equation (\ref{eq:const}): 

\begin{align}
    E_{\mathrm{flare},{\lambda}} = \int L_{\lambda,\mathrm{flare}} dt  = c \int L_\mathrm{flare} dt 
\end{align}

The calculated conversion constants for M1/M2 filters are given in Table \ref{tab:constant}.

\begin{table}[!ht]
      \caption{Conversion constants for the MuSCAT filters. }
         \label{tab:constant}
         \begin{center}
         \begin{tabular}{c c}
            \hline
            \noalign{\smallskip}
            \text{Filter}  & \text{Conversion constant} \\
            \hline
            \noalign{\smallskip}
            $g$ & 0.412 \\
            $r$ & 0.203 \\ 
            $i$ & 0.100  \\ 
            $z_\mathrm{s}$ & 0.058  \\
            \noalign{\smallskip}
            \hline
            \noalign{\smallskip}
         \end{tabular}
         \end{center}
     The constants are used to transform bolometric flare energies to passband-specific energies. We assumed here that the flare temperature is 7000 K.
\end{table}

\section{Flare temperatures}

\begin{table}[!ht]
      \caption{Ratio function for different MuSCAT filters and their asymptotic limits.}
         \label{tab:asymlim}
         \begin{center}
         \begin{tabular}{c c}
            \hline
            \noalign{\smallskip}
            \text{Ratio}      &  \text{Asymptotic limit}   \\
            \noalign{\smallskip}
            \hline
            \noalign{\smallskip}
            $R_{(g,r)}$ & 3.18  \\ 
            $R_{(g,i)}$ & 6.94 \\ 
            $R_{(g,z_\mathrm{s})}$ & 12.11  \\ 
%            \textit{R}_{(r,i)} & 2.18 \\ 
%            \textit{R}_{(r,z_s)} & 3.8  \\ 
%            \textit{R}_{(i,z_s)} & 1.74\\ 
            \noalign{\smallskip}
            \hline
            \noalign{\smallskip}
         \end{tabular}
         \end{center}
   \end{table}

\begin{table}[!ht]
      \caption{Peak S/N for both observed MuSCAT 2 flares.  }
         \label{tab:SNR}
         \begin{center}
         \begin{tabular}{c c c}
            \hline
            \noalign{\smallskip}
            & \hspace{10mm} \text{S/N} & \\
            \text{Filter}  & \text{flare 1} & \text{flare 2} \\
            \hline
            \noalign{\smallskip}
            $g$ & 17.5  & 4.3\\
            $r$ & 11.6 & 4.8\\ 
            $i$ & 5.4 & 3.76 \\ 
            $z_\mathrm{s}$ & 3.4 & 1.6 \\
            \noalign{\smallskip}
            \hline
            \noalign{\smallskip}
         \end{tabular}
        \end{center}
         
     There is a clear trend visible from higher S/N towards the bluer light for both flares.
\end{table}

\begin{figure}[!ht]
    \centering
    \includegraphics[width = \linewidth]{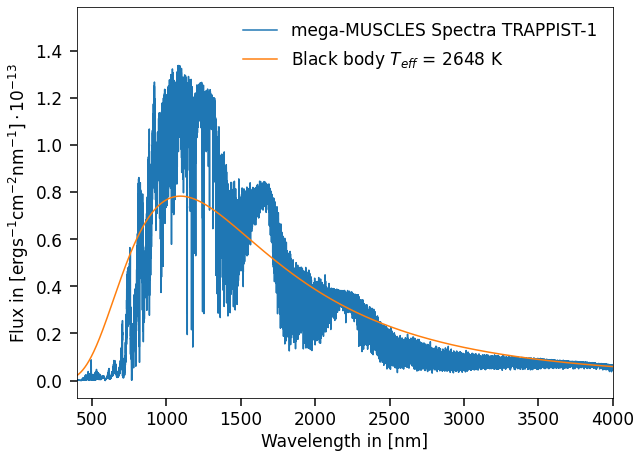}
    \caption{The semi-empirical mega-MUSCLES spectra for TRAPPIST-1 include emission and absorption lines. In comparison the black body temperature model for TRAPPIST-1 $T_\mathrm{eff} = 2648 $ K is shown. }
    \label{fig:megaMuscles}
\end{figure}

\begin{figure}[!ht]
    \centering
    \includegraphics[width = \linewidth]{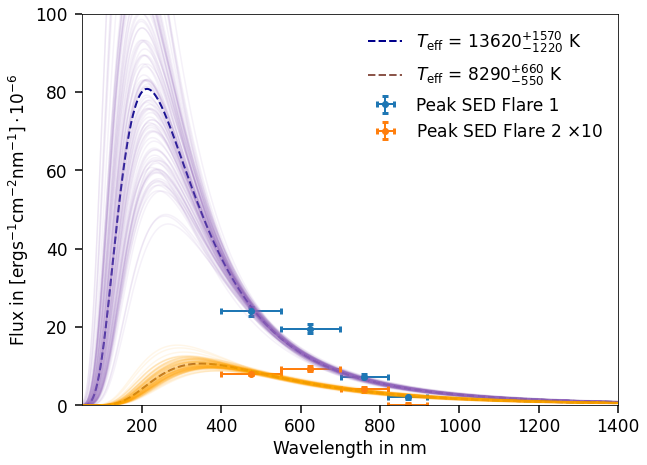}
    \caption{The SED of the peak fluxes for the two observed M2 flares. The SEDs fit with a model where the temperature $T_\mathrm{eff}$ was left as a free parameter using again \texttt{emcee}. The uncertainties of the model are indicated with 100 random samples from the chain.} 
    \label{fig:Peak_teff}
\end{figure}

\begin{figure}[!ht]
    \centering
    \includegraphics[width = \linewidth]{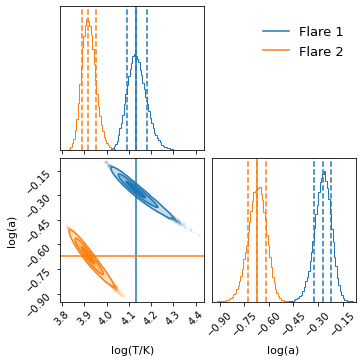}
    \caption{Posterior probability distribution for both peak flare samples: The sample of flare 1 is represented in blue and the samples of flare 2 are in orange. The sampling is done in logarithmic parameter space such that both the temperature $T$ and the area parameter $a$ are given on logarithmic scale. } 
    \label{fig:Peak_teff_post}
\end{figure}

%\subfigure{\includegraphics[width=0.24\textwidth]{figures/mega_muscles_TRAPPIST-1.png}} 
%

\begin{figure}[!ht]
    \centering
    \subfigure{\includegraphics[width=0.24\textwidth]{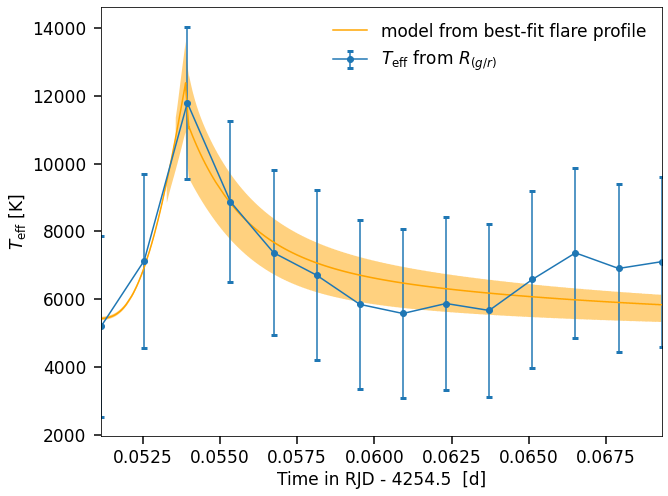}} 
    \subfigure{\includegraphics[width=0.24\textwidth]{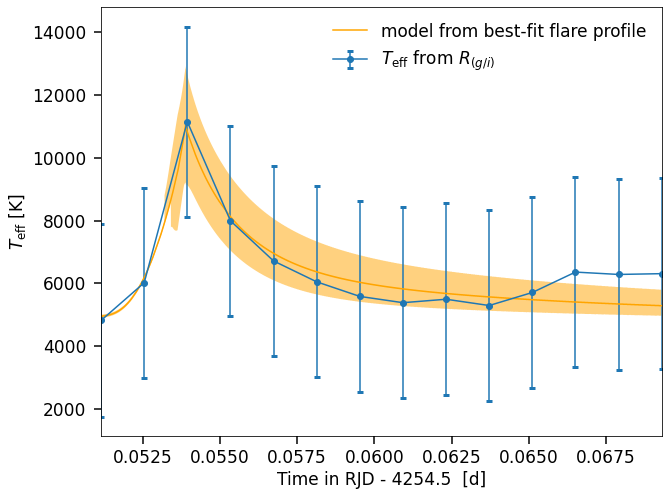}} 
    \subfigure{\includegraphics[width=0.24\textwidth]{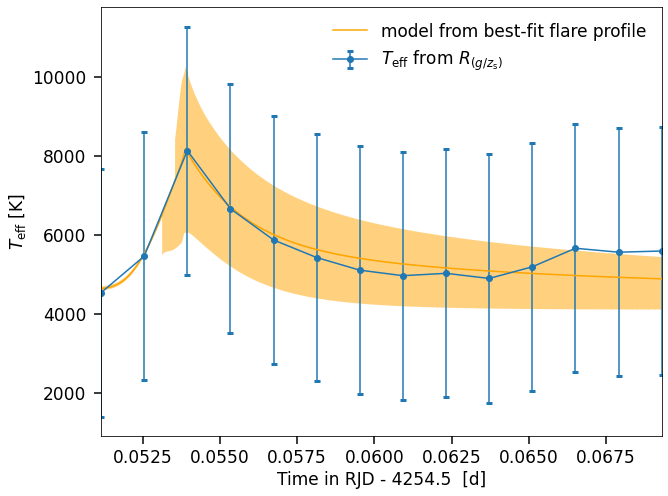}} 
    \subfigure{\includegraphics[width=0.24\textwidth]{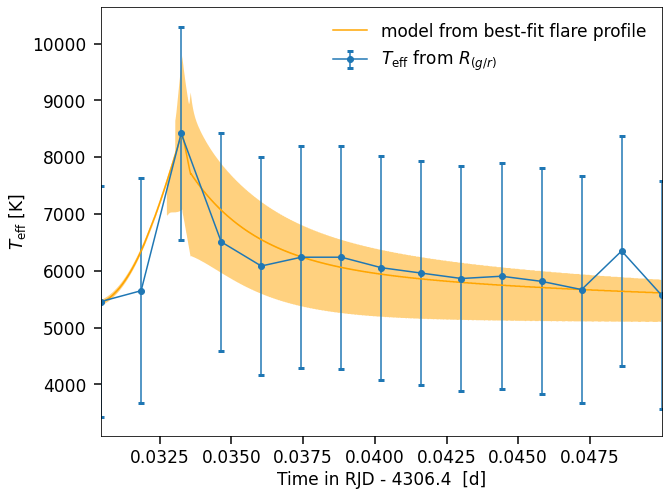}} 
    \subfigure{\includegraphics[width=0.24\textwidth]{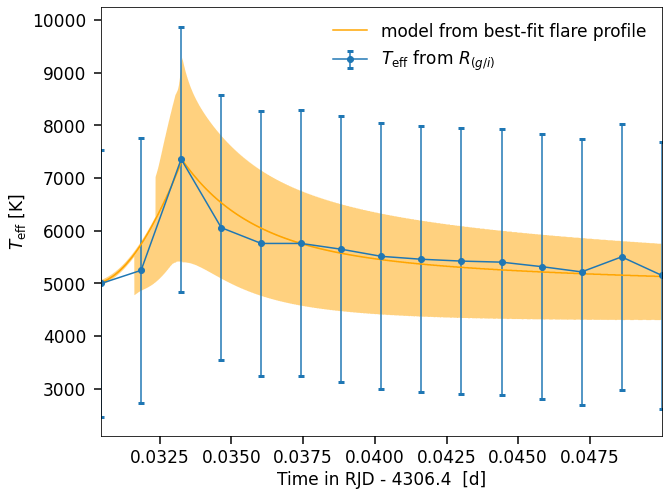}} 
    \subfigure{\includegraphics[width=0.24\textwidth]{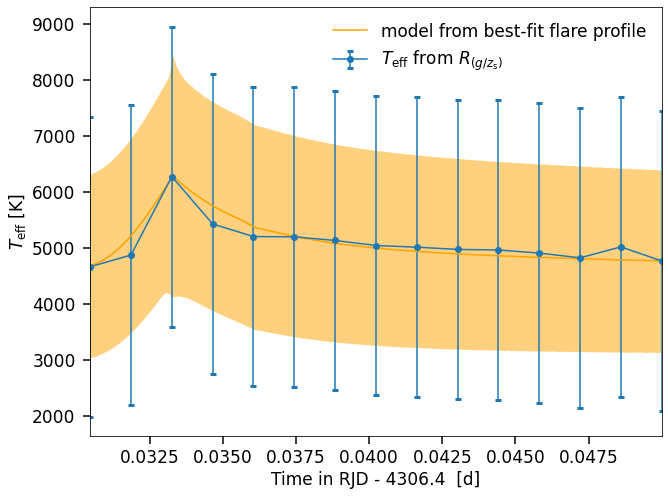}}
    \caption{Color-temperature evolution for both observed TRAPPIST-1 flare with MuSCAT 2. In the left column, all temperature evolutions for the different filter ratio functions $R_j(T_\mathrm{eff})$ are shown, and in the right column, the same is done for flare 2. The orange line indicates the model drawn from the best-fit to the flare flux profile, with respective model uncertainties represented by the orange shaded areas.  }
    \label{fig:temp_evolutio}
\end{figure}

\begin{figure}[!ht]
    \centering
    \includegraphics[width = \linewidth]{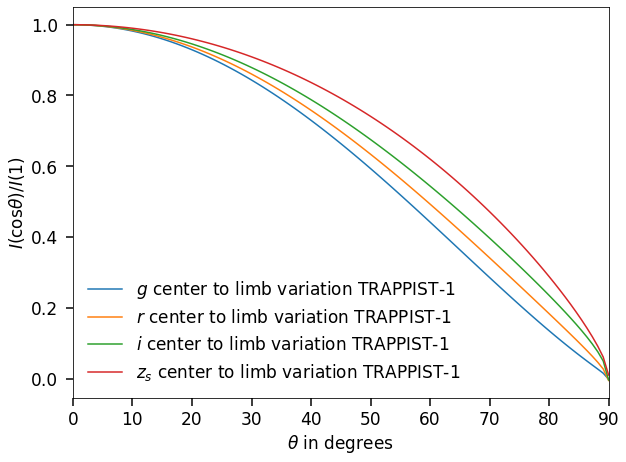}
    \caption{Center-to-limb variation for the MuSCAT filters. The limb darkening is modeled with quadratic law \citep{Claret_2000}. } 
    \label{fig:Center_2}
\end{figure}

%\begin{figure}
%    \centering
 %   \subfigure{\includegraphics[width=0.24\textwidth]{figures/MCMC_corner1.png}}
 %   \subfigure{\includegraphics[width=0.2225\textwidth]{f%igures/MCMC_corner.png}}
%    \caption{Histogram for both flares from the MCMC fit of the SEDs where the temperature is used as a free parameter. \hl{missing y-label, T [K]}}
    \label{fig:corner_flare}
%\end{figure}

\subsection{Flare black body model}
\label{subsec:model_BB}

In this section, we show our used black body model for the fitting process described in section \ref{subsec:Flare_Temperatures}. We start at the apparent brightness of the observer, 

\begin{align}
    l_\lambda = \frac{L_\lambda}{4\pi d^2},
\end{align}

with $L_\lambda$ the luminosity and $d$ the distance of the source. As the flare occurs on the visible side of the star, we integrate only the visible area from the star. Thus, we obtain

\begin{align}
    L_\lambda = 4\pi^2 R_*^2 B_\lambda(T_*),
\end{align}

with $R_*$  the radius of the star and $B_\lambda(T_*)$ the black body spectrum of temperature $T_*$. Substituting back in to the apparent brightness, 

\begin{align}
l_\lambda = \frac{R_*^2}{d^2} \cdot \frac{1}{\exp{\frac{hc}{k_b T_* \lambda}} - 1}  \cdot \frac{2hc^2}{\lambda^5} \cdot \pi.
\end{align} 

Under the assumption that a flare with an effective temperature of $T_\mathrm{flare}$ covers a circle with radius $R_\mathrm{flare}$ on the surface of the star, we can perform the same calculation. 

\begin{align}
l_{flare} = \frac{R_\mathrm{flare}^2}{d^2} \cdot B_\lambda(T_\mathrm{flare}) \cdot \pi = \frac{(a \cdot R_{*})^2}{d^2} \cdot B_\lambda(T_\mathrm{flare}) \cdot \pi \\ = \tilde{a} \cdot l_\lambda (T_\mathrm{flare}), \notag
\end{align} 

with $\tilde{a} = a^2$ and $a$ being the fraction of covered surface radius by the flare. In the optically thick case, we obtain the total apparent brightness, 

\begin{align}
    l_\mathrm{total} = \tilde{a} l_\lambda(T_\mathrm{flare}) + (1-\tilde{a}) \cdot l_\lambda(T_*).
\end{align}

As we obtain the apparent flare brightness directly as a result of our fitting pipeline, we fit the obtained brightness independent of the stellar brightness. Thus, we obtain the black body model for a flare with effective temperature $T_\mathrm{flare}$ and fractional area parameter $\tilde{a}$,  

\begin{align}
l_\mathrm{flare} = \tilde{a} \cdot (l_\lambda(T_\mathrm{flare}) - l_\lambda(T_*)).
\label{eq:bbmodel}
\end{align}

\begin{figure}[!ht]
    \centering
    \includegraphics[width = \linewidth]{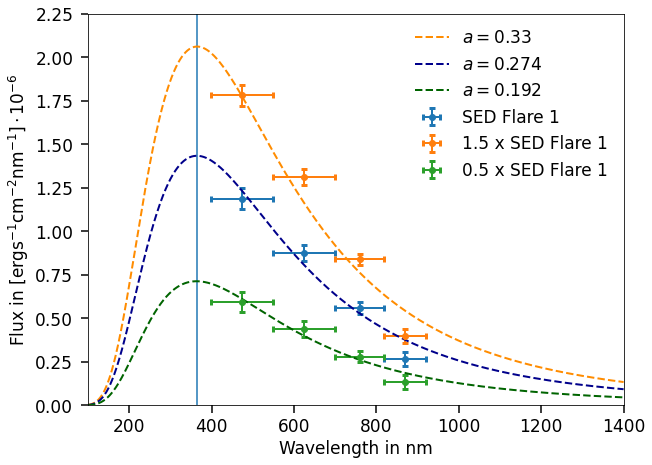}
    \caption{Model sensitivity test for the flare area parameter: three SEDs are shown for Flare 1. The blue SED indicates the SED with absolute calibration to Gaia and the orange and green curves represent fluxes with altered SEDs. } 
    \label{fig:model_sen}
\end{figure}

%In the optically thin case the subtraction needs to be replaced by an addition.  
\newpage
\end{appendix}

\end{document}